\documentstyle[epsf]{mn}
\def\epsfsize#1#2{\hsize}

\def\sgr{Sgr dSph }

\begin{document}

\title[The Sgr Dwarf Galaxy Survey I]
{The Sagittarius Dwarf Galaxy Survey (SDGS) I: 
CMDs, Reddening and Population Gradients.
First evidences of a very metal poor population.\thanks{Based on data 
taken at the New Technology Telescope - ESO, La Silla.}}

\author[M. Bellazzini et al.]
       {M. Bellazzini,$^1$
       F.R. Ferraro,$^1$
       and R. Buonanno,$^2$\\
        $^1$Osservatorio Astronomico di Bologna, Via Zamboni 33, 40125 
Bologna, ITALY\\
        $^2$Osservatorio Astronomico di Roma, Via dell'Osservatorio 2, 
00040, Monte Porzio Catone, Roma, ITALY}

\date{Accepted December 16, 1998;
      Received November 11, 1998;
      in original form July 7, 1998}
\pubyear{1998}

\maketitle

\begin{abstract}

We present first results of a large photometric survey devoted to study the 
star formation history of the Sagittarius dwarf spheroidal galaxy (Sgr dSph). 

Three wide strips (size $\simeq 9 \times 35 ~arcmin^2$) located at 
$\sim$($l^o;b^o$)=(6.5;-16), (6;-14), (5;-12) have been observed.
Each strip is roughly EW oriented, nearly along the major axis of the galaxy. 

A {\it control} field (size $\simeq 9 \times 24 ~arcmin^2$), located outside 
the body of \sgr [$\sim$($l^o;b^o$)=(354;-14)] has also been observed for 
statistic decontamination purposes. 

Accurate and well calibrated V,I photometry down to V$\sim 22$ have
been obtained for $\sim 90000$ stars toward the \sgr and $\sim 8000$ stars in 
the 
{\it control} field. 

This is the largest photometric sample (covering the widest spatial extension)
ever observed in the \sgr up to now.

The main new results presented in this paper are: (1) the possible discovery 
of a strong asymmetry in the distribution of stars along the major axis, since
the north-western arm of the Sgr galaxy (i.e. the region nearer to the Galactic
Bulge) apparently shows a significant deficiency of Sgr stars
and (2) the first direct detection of a very metal poor (and
presumably old) population in the Sgr stellar content. Hints for a metallicity
gradient toward the densest region of the galaxy are also reported.
 
\end{abstract}


\section{Introduction}

The \sgr is the most
prominent member of the family of dwarf spheroidals orbiting the Milky Way (MW)
both in luminosity ($\sim 10^7 L_{\odot}$) and extension on the sky
(at least $22 \times 8$ degrees). In addition, it is by far the  
galaxy nearest to the Sun [see Ibata et al. 1997 (IGWIS); Ibata, Gilmore \&
Irwin 1995 (IGI-II) and references therein].  

Since its discovery by
Ibata, Gilmore \& Irwin (1994, hereafter IGI-I), it was realized that the 
accurate study of this stellar system could help to shed light on many 
{\it hot topics}
of modern astrophysics, as for example the formation and evolution of
low surface brightness galaxies, the nature and extent of the dark matter
halo in dwarf galaxies and the phenomenon of accretion of galactic sub-units
by a main disc galaxy [see IGWIS and Montegriffo et al. (1998, hereafter MoAL) 
for 
discussions and references].
A huge observational and theoretical effort has been addressed to the study
of \sgr in the past three years and the research is still flourishing. 
We summarize here only the main observational results, the reader
interested to the theoretical simulations can find many useful references in
IGWIS, Hernandez \& Gilmore (1998), Zhao (1998), Ibata \& Geraint (1998) and 
Johnston (1998).   

>From a dynamical point of view, an estimate of the orbital period of the galaxy
has been obtained ($\sim 1 ~Gyr$) and strong cases have been made for the 
presence of a highly concentrated and massive dark matter halo (IGWIS,
Hernandez \& Gilmore 1998). The projected structure has been mapped over a
$\sim 15 \times 15 ~deg $ field by IGWIS, but many authors claimed the detection
of Sgr stars far away from the edges of the IGWIS isodensity map (Alard 1995,
Siegel et al. 1997). A huge project aimed to map the spatial structure of Sgr
using RR Lyrae stars is currently carried on by the OGLE
collaboration (Lepri et al. 1997).

H I observations (Koribalski, Johnston \& Otrupcek 1998) put a strong upper
limit on the gas content of the galaxy ($\sim 10^4 M_{\odot}$, i.e. less than
$10^{-3}$ of the total mass) confirming the classification of Sgr as a 
classical dwarf spheroidal galaxy (gas is generally missing in such galaxies,
see Koribalski, Johnston \& Otrupcek 1998, and references therein). 

Studies of the stellar content of Sgr has revealed the presence of carbon stars
(Whitelock, Irwin \& Catchpole 1996, hereafter WIC), planetary nebulae 
(Zjilstra \& Walsh 1996; Walsh et al. 1997) and RR Lyrae stars 
(Mateo et al. 1995a, 1995b; Alard 1996; Alcock et al. 1996). 
Color-Magnitude Diagrams (CMD) of Sgr stars have been
obtained by Mateo et al. (1995a; MUSKKK), Sarajedini \& Layden (1995; SL95) and
Marconi et al. (1997; MAL) while the properties of the globular
cluster system of the Sgr galaxy have been reviewed and discussed by MoAL.
Useful constraints about Sgr age and metallicity have also been found by 
Fahlman et al. (1996), Mateo et al. (1995b; hereafter MKSKKU) 
and Mateo et al. (1996).
The resulting scenario can be summarized as follows:
\begin{itemize}

\item The dominant stellar population (hereafter Pop A) of the \sgr is  
younger 
(by $\sim 4-5 ~Gyr$) than classical old galactic globulars (MUSKKK, MAL, 
Fahlman et al. 1996). MAL found that the main loci of Pop A CMD nearly coincide
with those of the young and metal rich globular cluster Ter 7 
(see Fusi Pecci et al 1995, and references therein), 
which belong to the Sgr globular cluster system (see MoAL). 
Average metallicity estimates 
ranges from $[Fe/H]=-1.1$ (MUSKKK) to $[Fe/H]\simeq -0.6$ 
(SL95). All authors, however, agree on the presence
of a metallicity spread, and the coexistence of at least two distinct
components
into the main population has been suggested by SL95. Preliminary results from
high resolution spectroscopy on a small sample of Sgr stars 
(Smecker-Hane , Mc William \& Ibata 1998) point toward an even larger 
spread in metal abundance, finding stars with metallicity from $[Fe/H]=-1.5$ to
$[Fe/H]=+0.11$. 

\item The simultaneous presence of a clear sequence of blue stars 
(Blue Plume, see below) and Carbon Stars have been interpreted as an
evidence for the presence of a sparse population of stars significantly younger
than Pop A (MUSKKK; WIC; Layden \& Sarajedini 1997, hereafter 
LS97). Let us call this component Pop B, for brevity. The estimates of
the absolute age of Pop B range from 4 Gyr (MUSKKK) to 1 Gyr (Ng 1997).

\item There are at least two (of the four) globular clusters associated 
with Sgr that are
very metal poor ($[Fe/H]\sim -2$) and significantly older than Pop A, 
i.e. M54 (MAL, LS97) and Terzan 8 (MoAL). 
This fact, coupled with the presence of RR Lyrae stars,
suggest the existence of an old and metal poor population in the galaxy.
However a direct identification of such a component is still missing. 

\end{itemize}

It is evident that many important pieces of information are still uncertain or 
missing and we are still far from having a satisfactory scenario for the star
formation and chemical enrichment history of Sgr. 
However, deriving clear-cut information from the CMDs of this galaxy 
is prevented by two fundamental factors:

\begin{enumerate}[rm]

\item \sgr is projected onto a region of the sky highly contaminated by
stars belonging to the central bulge and disc of the Milky Way;

\item the intrinsically low surface brightness ($\mu_V(0) \sim 25.5
~mag/arcsec^2$) coupled with the relatively small distance from the sun 
($\sim 25 ~Kpc$) implies that it is necessary to observe wide regions of the
galaxy to sample a significant number of member stars. The
sampling of stars in the post-MS phase is particularly
critical because of the short lifetimes, corresponding to low number densities
(see Renzini \& Fusi Pecci 1988).

\end{enumerate}

This latter point is fundamental in order to perform a
fruitful quantitative analysis of the Sgr stellar content. Assuming that
Sgr be a Simple Stellar Population\footnote{A conservative hypothesis in this
case, since at least two distinct populations are present.} 
(Renzini \& Fusi Pecci 1988) of age $\sim 10 ~Gyr$ and
$\mu_V(0) \sim 25.5~mag/arcsec^2$, $(m-M)_0=17.01$ and $A_V=0.55$ (MUSKKK), and
following the prescriptions of Renzini (1998) and Maraston (1998), the number
of stars in each evolutionary phase sampled by a given field of view can be
roughly estimated. Supposing, for instance, to image the densest region of 
Sgr with a field of view of $10 \times 10 ~arcmin$ we expect to sample only 
$\sim 300-400$ stars from the base to the top of the Red Giant Branch (RGB) 
and $\sim 50$ stars in the whole Horizontal Branch (HB). 
A HST-WFPC field would sample only $\sim 15$ RGB stars and $\sim 2$ HB stars. 

The former problem can be circumvented (at least partially) by careful 
statistical decontamination, using the CMD of stars representative of the 
contaminating
population (see MUSKKK as an example). However the success of any 
statistical decontamination depends on how well are defined the sequences
of the underlying population on the CMD. It is evident that both point (i)
and (ii) can be successfully afforded only sampling wide regions of the galaxy.

In this paper we present a photometric survey of the \sgr covering
a total field of view of $954 ~arcmin^2$.
 We obtained accurate V and I photometry for 89858 stars
(by far the largest CCD sample of Sgr stars to date), sampling three regions
of the galaxy $\sim 2^o$ apart from each other and located nearly along the
major axis ($l\sim 5^o$, see IGI-I, IGI-II and IGWIS). 

Two out of the three sampled regions included the fields recently surveyed with
HST (GO 6614, PI: K. Mighell) and for which very deep photometry (down to $\sim
4 ~mag$ below the MS-TO of Pop A) has been obtained (Mighell et al. 1997).
Since the limiting magnitude of the SDGS is V$\sim 22$ (just below
the Sgr Pop A Turn Off), the present study is the natural complement to the
deep (but ``small field'') photometry by Mighell et al. (1997).

Given the amount of collected data and the time needed to perform a complete 
and quantitative analysis, we decided to present the SDGS results in 
two papers. One considerable benefit of this approach is to provide the
astronomical community with the largest photometric database of Sgr stars 
and many useful guidelines for future observations {\em before} the full
completion of our analysis.   

The present paper (Pap I) is devoted to the general description of the
survey, and to set the basis for a detailed study of the Star Formation History
of the \sgr.   
 
We discuss data reductions, photometric calibration and completeness of the
samples (sect. 2). The CMD are shortly presented (sect. 3) and the differences 
in
interstellar reddening between the different fields are estimated (sect. 4).
Previous studies have shown that many useful things can be learned from Sgr
CMDs before (and/or without) performing statistical decontamination of the
whole diagrams (MUSKKK, SL95, MAL). We include here also many tests devised 
to study the Sgr stellar populations and their projected 
spatial distribution, either using the ``raw'' (foreground contaminated) 
CMDs (sect. 5 and sect. 6) or attempting to remove the contribution of 
foreground stars from star counts in small selected box on CMDs (see sect. 5.5).
The derived results and suggestions will be used as a guideline and 
independent check for the forecoming analysis. 
A final summary is reported in sect. 7.

The companion paper (Bellazzini et al. 1998b; Pap II, in preparation) will be 
devoted to the detailed analysis of statistical decontaminated CMDs, 
specifically aimed to the study of the Star Formation History in Sgr.  

\section{Description of the SDGS}

The main purpose of the SDGS was to sample a statistically significant number of 
evolved stars in the \sgr. A wide area of sky was covered
using a ``snapshot survey technique'', i.e. taking relatively
short (V,I) exposures (long enough to reach the TO level) in many different
fields using a camera with a wide field of view. Two nights (27 - 28, June 
1997) at the ESO 3.5 m New Technology Telescope (NTT) were allocated to this
project .

We used the EMMI camera with the RILD setup (Zjilstra et al. 1996).
This instrument provides an optically corrected
and unvignetted field of $9.15\times 8.6 ~arcmin$. The detector was a Tek
CCD ($2048\times 2048$ pxs, pixel size $ 24 \mu m$) with a read-out
noise of $3.5 e^-$ rms, a dark current of $3 e^- /hr$ and a gain factor
of 1.34 $e^-/ADU$. The image scale is $0.27 ~arcsec/px$. The filters used are
ESO 606 (V) and ESO 610 (I). In the best cases we acquired a {\em long} (300 s) 
and a {\em short} (60 s) exposure in each filter.

We observed three strips (nearly E-W oriented, i.e. approximately along the Sgr
major axis) each composed by five partially overlapping fields. 
An overlapping area of nearly $1.5 ~arcmin$ has been secured between each
couple of adjacent fields in order to obtain a robust relative photometric 
calibration in each strip.

As already mentioned , the location of two strips has been chosen to include the
fields covered by deep HST observations:  

\begin{figure*}
 \vspace{20pt}
\epsffile{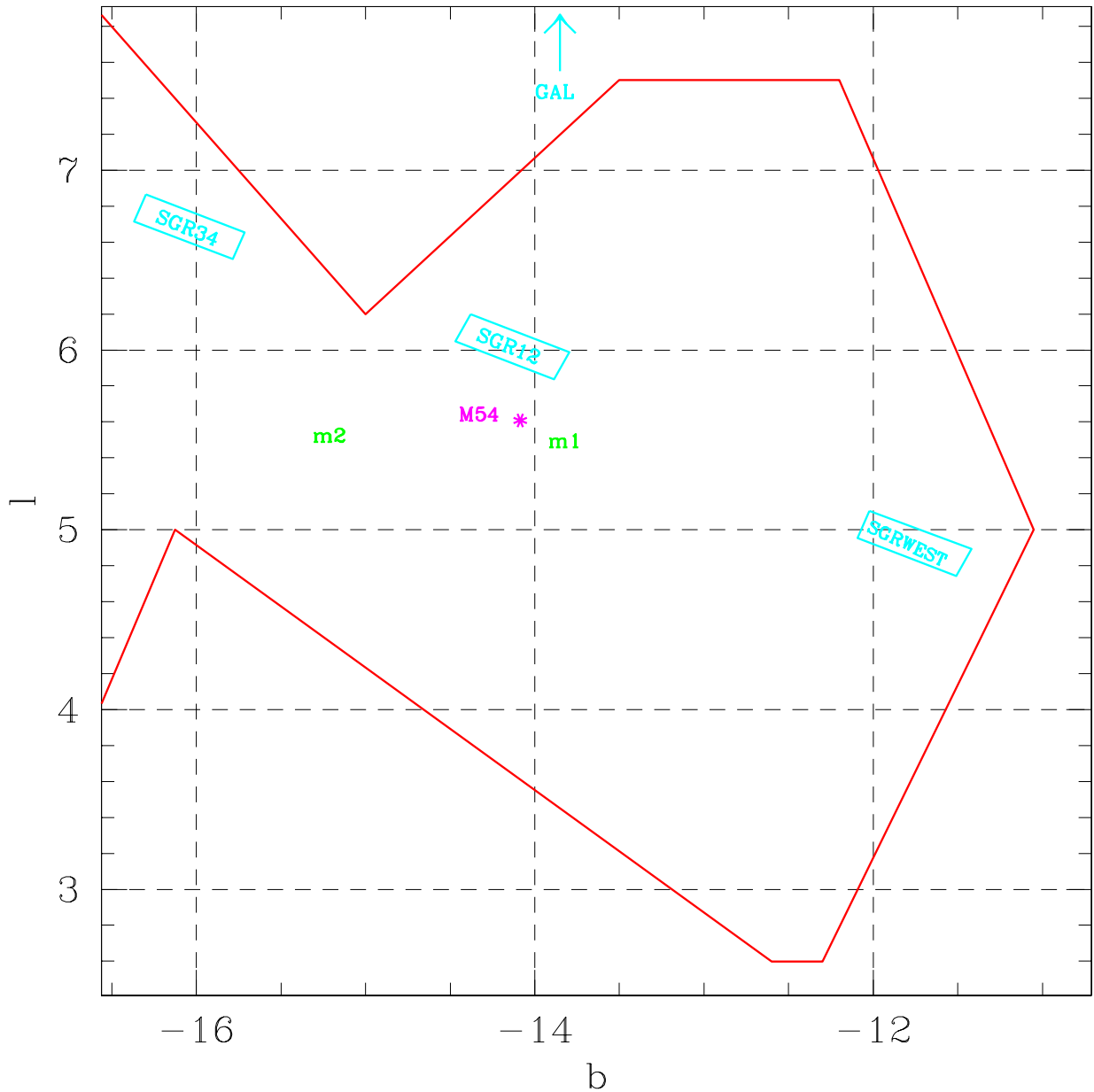}
 \caption{A map in galactic coordinates representing the position of the
 observed fields within the Sgr galaxy. The heavy line roughly reports
 the limit of the outer isodensity contour of the IGWIS map (their fig. 1).
 The position of SGR34, SGR12 and SGRWEST strips is 
 reported; the arrow at the top of the panel indicates the direction
 toward the control GAL field. The position of the M54 globular cluster is
 indicated by an asterisk and the location of the two fields observed by MAL
 is indicated by the labels m1 and m2. The field observed by MUSKKK is included
 in SGR34, the Sgr field observed by SL95 is included in SGR12. The small fields
 observed with HST by Mighell et al. (1997) are included in SGR34 and SGR12.}
\end{figure*}

A first strip, located near the center of density 
of Sgr (IGWIS) and extending from ($l=5.8^o;b=-13.9^o$) to 
($l=6.0^o;b=-14.4^o$), 
includes the HST fields labeled Sgr 1 ($\alpha_{2000}= 18~h ~55~m
~;~ \delta_{2000}= -30^o ~16^{\prime}$; also observed by SL95) and Sgr 2 
($\alpha_{2000}= 18~h ~55~m ~;~ \delta_{2000}= -30^o ~14^{\prime}$); hereafter 
we will call it SGR12 strip.

A second strip, located $\sim 2^o$ eastward of SGR12
and extending from ($l=6.6^o;b=-16.2^o$) to ($l=6.5^o;b=-15.8^o$), 
include the HST fields labeled Sgr 3 ($\alpha_{2000}= 19~h ~06~m
~;~ \delta_{2000}= -30^o ~25^{\prime}$) and Sgr4 
($\alpha_{2000}= 19~h ~06~m ~;~ \delta_{2000}= -30^o ~23^{\prime}$; both fields
were also observed by MUSKKK); hereafter we will call it SGR34 strip.

A third strip was observed in a still ``unexplored'' region of Sgr, westward
of SGR12, from ($l=4.9^o;b=-12.1^o$) to ($l=4.7^o;b=-11.5^o$); hereafter
we refer to this strip as SGRWEST, since the sampled region can be considered
as the north-western arm of the Sgr galaxy. This is the field observed at 
the lowest galactic latitude: it was selected to study the Sgr population in a 
region where the interaction with the Galaxy should be stronger 
(according to IGWIS, this is the region of the \sgr that is leading the galaxy
along its orbit, and so will be the first one to ``collide'' with the Galaxy
during the present perigalactic passage).
It is important to note that SGR34 and SGRWEST fields are nearly symmetrically
located with respect to the center of density of the \sgr (SGR12), 
along the major axis: SGRWEST is $2^o$ from the center in the direction of the 
Galactic Bulge, while SGR34 is $1.6^o$ in the opposite direction. 

In addition we observed a control field in a region devoid of Sgr stars and
dominated by foreground Galactic field stars. This field has been observed with
the same ``strip technique'' described above, but only three - partially
overlapping - EMMI fields have been observed in this case. 
The Galactic field strip include the control field labeled SGR-MWG1 and
SGR-MWG2 in the HST-GO 6614 proposal and also observed by MUSKKK, around
($\alpha_{2000}= 18~h 33~m ~;~ \delta_{2000}= -40^o ~56^{\prime}$); we will
call this strip the GAL field.

Bad weather conditions and some technical problem occurred during the allocated
nights prevented us from observing two additional strips 
(located between SGR12 and SGR34) which were originally planned to 
complete the SDGS.

The location (in Galactic Coordinates) and approximate dimensions of 
the fields covered by the SDGS are presented in fig. 1. The heavy line roughly
represent the most external isodensity line of the IGWIS map (their fig. 1).
The position of the fields observed by MAL is also reported. The arrow indicates
the direction toward the GAL field which lies
outside the limits of the figure but is located at the same Galactic latitude 
of SGR12. 
 
\subsection{Observations}

The log of the observations is presented in Table 1.
A grand total of 58 frames has been obtained.

Each field within a given strip has been named A, B, C, etc., in sequence,
assigning label A to the frame used as reference frame for the relative 
calibration within the strip (see sec. 2.2). 
The columns of Table 1 reports, in order,
the name of the observed field, galactic longitude ($l^o$) and latitude ($b^o$)
of the center of the field, the adopted filter, the exposure times
and the seeing conditions (FWHM of the Point Spread
Function). 

\begin{table}
 \centering
 \begin{minipage}{140mm}
  \caption{Observations report.}
  \begin{tabular}{@{}lccccc@{}}
Field & $l^o$ & $b^o$ & Filter  & Exp.  & Seeing \\
Name  &       &       &        &  time &  (FWHM) \\
&&&&&\\
SGR12 A & 5.99 & -14.36 & V & 300,  60 s & $1.92^{''}$\\
SGR12 A & 5.99 & -14.36 & I & 300,  60 s & $1.71^{''}$\\
SGR12 B & 5.95 & -14.27 & V & 300,  60 s & $1.90^{''}$\\
SGR12 B & 5.95 & -14.27 & I & 300,  60 s & $1.57^{''}$\\
SGR12 C & 5.91 & -14.17 & V & 300,  60 s & $1.87^{''}$\\
SGR12 C & 5.91 & -14.17 & I & 300,  60 s & $1.51^{''}$\\
SGR12 D & 5.87 & -14.08 & V & 300,  60 s & $2.23^{''}$\\
SGR12 D & 5.87 & -14.08 & I & 300,  60 s & $1.40^{''}$\\
SGR12 E & 5.83 & -13.99 & V & 300,  60 s & $2.81^{''}$\\
SGR12 E & 5.83 & -13.99 & I & 300,  60 s & $1.46^{''}$\\
&&&&&\\
SGR34 A & 6.66 & -16.26 & V & 300      s & $1.24^{''}$\\
SGR34 A & 6.66 & -16.26 & I & 300      s & $1.34^{''}$\\
SGR34 B & 6.62 & -16.17 & V & 300      s & $1.28^{''}$\\
SGR34 B & 6.62 & -16.17 & I & 300      s & $1.22^{''}$\\
SGR34 C & 6.58 & -16.07 & V & 300,  60 s & $1.59^{''}$\\
SGR34 C & 6.58 & -16.07 & I & 300,  60 s & $1.34^{''}$\\
SGR34 D & 6.54 & -15.98 & V & 300,  60 s & $1.43^{''}$\\
SGR34 D & 6.54 & -15.98 & I & 300,  60 s & $1.46^{''}$\\
SGR34 E & 6.50 & -15.89 & V & 300,  60 s & $1.34^{''}$\\
SGR34 E & 6.50 & -15.89 & I & 300,  60 s & $1.41^{''}$\\
&&&&&\\
SGRWEST A & 4.92 & -12.01 & V & 300       s & $1.12^{''}$\\
SGRWEST A & 4.92 & -12.01 & I & 300       s & $1.22^{''}$\\
SGRWEST B & 4.87 & -11.91 & V & 300       s & $1.20^{''}$\\
SGRWEST B & 4.87 & -11.91 & I & 300       s & $1.40^{''}$\\
SGRWEST C & 4.82 & -11.80 & V & 300       s & $1.20^{''}$\\
SGRWEST C & 4.82 & -11.80 & I & 300       s & $1.16^{''}$\\
SGRWEST D & 4.78 & -11.72 & V & 300       s & $1.18^{''}$\\
SGRWEST D & 4.78 & -11.72 & I & 300       s & $1.37^{''}$\\
SGRWEST E & 4.72 & -11.59 & V & 300       s & $1.16^{''}$\\
SGRWEST E & 4.72 & -11.59 & I & 300       s & $1.12^{''}$\\
&&&&&\\
GAL A & 353.88 & -14.25 & V & 300,  60 s & $1.43^{''}$\\
GAL A & 353.88 & -14.25 & I & 300,  60 s & $2.12^{''}$\\
GAL B & 353.77 & -14.30 & V & 300,  60 s & $1.46^{''}$\\
GAL B & 353.77 & -14.30 & I & 300,  60 s & $1.95^{''}$\\
GAL C & 353.67 & -14.35 & V & 300,  60 s & $1.50^{''}$\\
GAL C & 353.67 & -14.35 & I & 300,  60 s & $1.38^{''}$\\
\end{tabular}
\end{minipage}
\end{table}

As shown in Tab. 1, the seeing conditions were never particularly good but 
were remarkably stable when the SGR34 and SGRWEST fields were observed. 

On the contrary, significant seeing
variations has been encountered during the acquisition of the SGR12 frames. 
The net result was a brighter limiting magnitude for SGR12 frames
with respect to SGR34 and SGRWEST, and a bad quality Color Magnitude Diagram 
below $V\sim 19$ for the fields SGR12 D and E (only stars with $V\le 18.5$ from
these two fields will be included in the final CMD of the SGR12 strip, see
below). 

GAL frames were obtained during the moon rising phase (nearly at the end of the
night), and show a high average sky level, resulting in a brighter
limiting magnitude with respect to SGR34 and SGRWEST.

Short exposure frames are missing for SGR34 A and B fields and for the whole
SGRWEST strip. This is not a serious problem for our purposes, since the RGB
tip of the Sgr Pop A is located at $V\sim 16$ while the saturation level of the
SDGS long exposures occurs at $V\sim 15.5$, even in the best seeing conditions.
Thus we are confident that even the images of the brightest Pop A giants are 
not significantly saturated; however their position in the CMD may be affected 
by larger uncertainties (with respect to the cases when also short exposures 
were secured) since they could have been observed in the non-linear regime of 
the CCD response.

\subsection{Data reduction}

All the frames were trimmed, bias subtracted and flat-fielded, with high S/N
flat field frames secured at the beginning and the end of the nights.
The pre-reduction procedures were done using the MIDAS package. 

The photometric reductions were carried out using the ROMAFOT package 
(Buonanno et al. 1983, and references therein) mounted on a Digital-Alpha 
station of the Bologna Observatory.
A standard two-dimensional PSF fitting procedure was performed on the
pre-reduced frames.
In order to determine the PSF and to find and fit stars in each frame,
we adopted the standard procedure as described, for example, 
by Ferraro et al. (1990). The PSF parameters were allowed to vary as a function
of position in the frame, to account for slight differences in the images shape 
across the wide EMMI field of view. 

The program stars were detected automatically adopting a threshold-criterion
($\sim 5 \sigma$) above the local background level, for the deep exposures,
and ($\sim 6 \sigma$) for the short ones.
The searching procedure was performed on the V frames, and the list of 
identified sources were used as input for the PSF fitting in the corresponding 
I frames.

The instrumental magnitudes of each frame in a given strip were referred to 
the instrumental magnitudes of the frame assumed as a reference (i.e. A fields). 
Stars in common in the overlapping region (typically $\sim 1500$ for each couple
of adjacent fields) were used to determine the transformations between
instrumental magnitudes of a given frame and the reference
frame. The transformations were always well defined and never larger than 
few hundredths of magnitude.
The uncertainty in the relative frame-to-frame calibration within a strip 
is $\sim 0.01 ~mag$ (conservative estimate). 

\begin{figure*}
\vspace{20pt}
 \caption{Internal photometric errors in each observed strip. The magnitude
 differences for stars in common between two adjacent fields is plotted versus
 the V mean magnitude. In all cases the photometric uncertainty is 
 $< 0.05 mag$ (dashed line) for stars brighter than $V = 19$.}  
\end{figure*}

\subsection{Photometric errors}

The best empirical estimate of the internal accuracy of a photometric data set
is usually derived from the rms of repeated measures of selected stars. 
We used stars in the overlapping regions: the magnitude differences for these
stars as a function of V magnitude are plotted in fig. 2.
We assume the {\em standard deviation} around zero (computed in bins 1 mag wide) 
of the magnitude differences as a conservative estimate of the
{\em average} photometric uncertainty in the corresponding magnitude bin.
For each considered strip, the trend of the photometric error with V magnitude 
was optimally fitted with a polynomial, and so it has been possible to
associate a typical photometric error to each star, as a function of its V
magnitude. As can be seen from fig. 2, the internal accuracy is better than
0.05 mag for the large majority of the measured stars.
  
\subsection{Absolute calibration}

Several CCD standard fields (Landolt 1993) were acquired during both the nights
in order to provide an accurate absolute calibration. A total of 23 standard
stars, covering a wide range of colors, were used to link the instrumental
aperture magnitudes to the standard Johnson system.

The reduction of
standard star observations and the determination of aperture correction for
the science frames has been performed with a standard technique, using the
appropriate task of ROMAFOT devoted to aperture photometry.

In fig. 3, the differences between standard (V,I) and 
instrumental (v,i) aperture magnitudes vs. instrumental color index 
are reported. 
Since the slope is negligible with respect to the 
dispersion around the mean we decided to exclude color terms in the adopted 
calibration equations. The final adopted equations are the following:

$$V = v + 25.21 \pm 0.02$$
$$I = i + 24.61 \pm 0.02$$ 

where the error is a conservative estimate of the uncertainty in the absolute
calibration.

A set of most isolated, unsaturated stars in the reference frames were 
then used to link the ``aperture'' magnitudes to the PSF-fitting 
instrumental magnitudes. The ``fit-to-aperture'' corrections applied introduces
an extra zero-point uncertainty of $\sim 0.02$ mag in each color and has to be
regarded as the major source of uncertainty in the absolute calibration of 
the stars in each strip of SDGS. 

\begin{figure}
 \vspace{20pt}
\epsffile{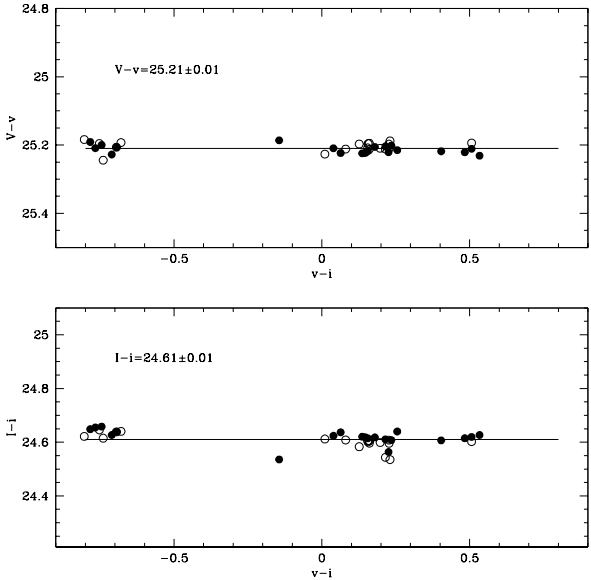}
 \caption{Calibration curves as a function of instrumental (v-i) color for the
 observed Landolt stars. Different symbols refer to stars observed during
 different nights: June, 27 (open circles) and June, 28 (filled circles),
 respectively.}  
\end{figure}

The calibrated data are available via anonymous FTP at hopi.bo.astro.it in the
directory /home/ftp/pub/michele/SDGS-I. 

\subsection{Completeness of the samples}

The crowding conditions in the observed fields are not critical ($ 0.03
~stars/arcsec^2$ in the worst case) and are very
homogeneous in general. This suggests that the completeness level should be not
very different from field to field, at least within a given strip.
Some preliminary artificial star test confirmed this expectation.
Thus we decided to perform extensive artificial star experiments in {\em one
field for each strip}, assumed to be representative of the whole strip
sample, namely SGR12 B, SGR34 C, SGRWEST B and GAL C.

A total of nearly 4000 artificial test stars were added to each selected field,
never more than 100 each time, in order to leave unaltered the crowding 
conditions.
The test stars were chosen with a color representative of the mean
color of the sampled population. A square sub-image containing the selected
star were extracted from the original frame and randomly added to the frame.
The entire reduction procedure has been repeated in the same way on the
synthetically enriched images. The ratio between the number of simulated stars
recovered with a magnitude within $\pm 0.1 ~mag$ from the 
original test star, and the total number of artificial images added was taken
as the completeness factor ($C_f$) in the corresponding magnitude bin. The
typical uncertainties on the completeness factors is $\le 5 \%$.

\begin{figure}
 \vspace{20pt}
\epsffile{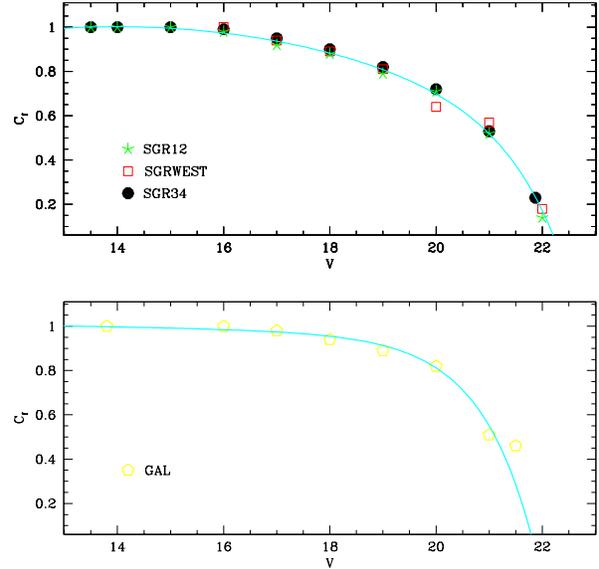}
 \caption{Completeness factor as a function of V magnitude for the SGR12 B, 
 SGR34 C and SGRWEST B fields (upper panel) and for the GAL C field. The
 reported lines are best fit curves to the data. The resulting completeness
 functions is adopted as representative of the whole sample in a strip:
 the curve in the upper panel is used for SGR12, SGR34 and SGRWEST (since
 the differences among the measured completeness factors are negligible); the
 curve in the lower panel is used for the GAL strip. The uncertainty on each
 completeness factor is $\le 5 \%$.}  
\end{figure}

The results of the completeness experiments for SGR12 B, SGR34 C and SGRWEST B
are reported in the upper panel of fig. 4. It is evident that the completeness
factors for this three fields are the same within the errors 
in the common range of magnitude ($16 < V < 21$). This is not
unexpected since the crowding conditions are nearly similar and in any case
very far from the critical conditions found, for instance, in the central
region of globular cluster images. Furthermore the (x,y) distribution of stars 
is very homogeneous so there is no expected variation of the completeness with
position. This results provide a robust support to use the derived 
completeness as representative of the whole considered
strips, since it has been demonstrated that the observed field to field 
variations have a minimal impact on the overall completeness.

We decided to adopt a unique completeness function for the three quoted strip,
which is plotted as a solid curve in figure 4 (upper panel). 
Note that $C_f$ falls below 0.5 only at $V\simeq 21$ and is $> 0.8$ 
for $V\le 19$.

A distinct completeness function is adopted for GAL (fig. 4, lower panel),
since little but significant differences are present, with respect to the 
Sgr strips. 
Higher values of $C_f$ are found in the range ($18 < V < 20$), probably
because of the significantly lower stellar density (i.e. better crowding
conditions).

If not otherwise stated, any star count will be hereafter corrected for
completeness using the completeness functions presented in fig. 4, in the range
from the brighter unsaturated stars to the level where $C_f$ fall below 0.5.

A useful by-product of the artificial stars experiments, is an independent check
of the ability of the photometry code to reproduce the same result in repeated
measures of the same stars. The standard deviation of the difference
between the original magnitude of a given star and the derived magnitude of
the recovered copies artificially distributed in the frame was found to be 
always significantly less than the adopted total photometric error described 
in sect. 2.3. 
  
\section{Color Magnitude Diagrams}

\subsection{Homogeneity of the samples}

Following the procedure described in sec. 2.2, a CMD for each EMMI field has
been obtained. The single-field CMDs in each strip have been compared in order
to check for possible systematic differences in reddening (see below)
and stellar content within each strip. 
Moreover, specific population ratio tests in selected region
of the CMDs were performed to obtain quantitative intercomparisons.
No significant difference in the distribution of stars in the CMD emerged from
these tests, the stellar content within each strip turned out to be very
homogeneous. For this reason we decided to merge the sample of each field
of any given strip in one single database for each strip. 
For the stars lying in the overlapping regions, the average magnitude between
the two available measures has been adopted.

Visual inspection of single-field CMDs also suggested that the possible impact
of differential reddening within each strip must be negligible.

 A further check was made possible by
the availability of the new DIRBE/IRAS All-Sky Reddening Maps (ASRM; Schlegel,
Finkbeiner \& Davis 1997, 1998) which allow reddening estimates with a spatial
resolution of $\sim 6 ~arcmin$ and a typical accuracy of $\sim 16 \%$, so
superseding the widely used maps by Burstein \& Heiles (1982). 

We calculated from ASRM (see Stanek 1998, for details of the procedure) the
E(V-I) in the center position of each EMMI single field interpolating from four 
adjacent
resolution elements of the map. The derived reddening differences among 
fields in each strip are always small, the maximum difference being less than
0.02 mag. 
The standard deviation about the mean E(V-I) value of
single fields [$\sigma_{E(V-I)}]$ amounts to 0.01 mag for SGR34
and SGRWEST, to 0.015 for GAL and to 0.005 for SGR12 [see also table 2, below;
we adopted all over the paper
the following relations: $E(V-I)=1.6\times E(B-V)$ and $A_V=1.93 \times
E(V-I)$; from Rieke \& Lebovsky 1985, hereafter RL].
With this further support, we conclude that any possible differential 
reddening between single fields is of the same order of the uncertainty of the 
relative calibration or less, and so negligible for the present purposes.

In conclusion we can state that, at the present level of accuracy, the stars
in each strip can be considered as an homogeneous sample from any point of
view. 

Hereafter, any distinction between sub-fields within a strip is therefore
useless and we will refer to the whole strip samples with the labels SRG12, 
SGR34, SGRWEST and GAL {\em Fields} (capital letter). 
We will refer to the sample of stars from the SGR12 A,
B and C fields alone with the label SGR12R (Restricted). 

\subsection{The global CMDs}

In figures 5, 6, 7 and 8 the total CMDs of the three Sgr strips and of the
control field are displayed. In fig. 5, only the stars belonging to SGR12 A, B
and C fields are reported since, as already mentioned, the faint part of 
SGR12 D and E CMDs have poor photometric quality. 

\begin{figure*}
 \vspace{20pt}
 \caption{Color Magnitude diagram for all the stars measured in the A, B, C
 fields of the SGR12 strip (16992 stars).}  
\end{figure*}

Here we make only few comments on the CMDs, mainly comparing the Sgr  
CMDs to the control field one. Deeper insights will be added in
sec. 5 after the discussion of reddening. 

The main general features of the CMDs are the following:

\begin{enumerate}

\item apart from the different photometric quality, 
SGR34 (fig. 6) and SGR12 (fig. 5) diagrams are very similar and show 
all the characteristics of the typical Sgr CMDs (MUSKKK, SL95, MAL). 
The bulge+disc contaminating population 
(compare with the control filed CMD in fig. 8) produce a pronounced 
{\em blue sequence} at $V-I\sim 0.8$, that is nearly vertical from 
$V\sim 15$ to 
$V\sim 20$ where it become a large band sloped toward redder colors. 
A sparser feature runs near parallel to the vertical part of this sequence, 
at slightly redder color ($V-I\sim 0.9 ~-~0.1$), and can be identified with 
the giant branch of the bulge population.

\item The Sgr population is clearly evident as a {\em globular cluster-like} 
sequence with a Main Sequence Turn Off (MSTO) around ($V\sim 21.5; V-I\sim0.7$), 
a well defined (at least in the SGR34 CMD) Sub Giant Branch (SGB),a wide
Red Giant Branch (RGB) extending to ($V\sim 16; V-I\sim 2.5$) and a clear red
Horizontal Branch (HB) at ($V\sim 18 - 18.3; V-I\sim 1$). The SGR12 HB appears
more clumped to the red with respect of that of SGR34. In the SGR34 and SGR12
fields this globular cluster-like population is clearly the dominant feature
of the CMDs (Pop A). 

The blue plume discussed by MUSKKK, SL95, Ng (1997), LS97, and identified with
a young component (Pop B, see above) is clearly present at 
($V\sim 20; V-I\sim 0.4$)

\begin{figure*}
 \vspace{20pt}
 \caption{Color Magnitude diagram for all the stars measured in the SGR34 
 strip (22603 stars).}  
\end{figure*}

\item The SGRWEST CMD appears dominated by the contaminating population, in
particular the bulge MS. The MSTO of the bulge seems to emerge from the 
vertical blue sequence at ($V\sim 18 - 19; V-I\sim 0.8$). However the
signatures of the Sgr population are still present, in particular the sparse
RGB plume and the red HB. The Sgr TO and SGB are not well defined but 
their presence is strongly suggested by a clear excess of stars in the region
around $V\sim 21.5$ and $V-I \sim 0.6-1.0$, with respect to the GAL CMD.

\end{enumerate}  

\begin{figure*}
 \vspace{20pt}
 \caption{Color Magnitude diagram for all the stars measured in the SGRWEST 
 strip (41462 stars).}  
\end{figure*}

The comparison of the CMDs in fig. 5, 6, 7 with the corresponding ones 
derived by other authors (MUSKKK for SGR34 and GAL; SL95 for SGR12; MAL) 
shows, in general, an excellent agreement. 
The positions of the main loci are compatible within
the photometric errors and any possible disagreement in
the absolute calibration can be considered negligible. The most robust
comparison can be performed between SGR34 and the MUSKKK CMDs, since they are
similar both in limiting magnitude level and in overall photometric quality.
The agreement is very good. In Pap II we will take advantage of the deeper
control field CMD obtained by MUSKKK at the same position of the GAL strip to 
properly decontaminate the TO region of our Sgr CMDs. 

\begin{figure*}
 \vspace{20pt}
 \caption{Color Magnitude diagram for all the stars measured in the control
 field, the GAL strip (8336 stars).}  
\end{figure*}

\section{Reddening}

Once stated the homogeneity of the samples {\em within} Fields we can turn to 
study the possible differences {\em among} the observed Fields. Before
discussing the stellar content of the Fields it is
advisable to check if the SGR12, SGR34, SGRWEST and GAL Fields are affected 
by different amount of interstellar extinction and reddening.

\subsection{Differences between SDGS Fields}

At present, there are two direct reddening estimates in Sgr regions sampled by
SDGS, i.e. $E(V-I)=0.22\pm 0.04$ for SGR34 (MUSKKK) and $E(B-V)=0.13\pm 0.02$
for SGR12 (SL95), which corresponds to $E(V-I)=0.21\pm 0.03$ (RL). Thus the two 
measures agrees within the errors (but see SL95 for further discussions). 
No estimate exist for SGRWEST
and GAL Fields, however MUSKKK implicitly assume that their Sgr field (a
subsample of SGR34) and their control field (a subsample of GAL) have the same
extinction.

The reddening measure by MUSKKK must be
regarded as the most robust one, since it has been directly derived from the 
colors at minimum of 5 Sgr RR Lyrae stars. On the other hand SL95 used the SMR
method (Sarajedini 1994, hereafter S94) devoted to simultaneously derive 
the metallicity and reddening from the observed RGB ridge line and 
Horizontal Branch luminosity ($V_{HB}$) of globular clusters (GC). 
In the present case the power of the method can
be seriously weakened by the intrinsic wideness of the RGB, which on the
contrary is usually very thin and well defined for GCs. Furthermore, Sgr
population is almost certainly younger - and perhaps slightly more metal rich -
than the Galactic globulars used to calibrate the SRM method. 
We plan to test the application of the 
SMR method to SDGS data only on decontaminated CMDs (Pap II). 

Thus, we assume $E(V-I)=0.22\pm 0.04$ (MUSKKK) for 
SGR34 as an ``absolute reddening zero-point'' and try to determine eventual 
reddening difference between SGR34 and the others SDGS Fields by other means.

A first check has been performed using ASRM, as described in the previous
section. It result (see tab. 2, below) that SGR12, SGR34 and SGRWEST have very
similar reddening, and the absolute value is in excellent agreement with the
direct estimates ($E(V-I)\sim 0.21 ~- ~0.22$). 
A lower reddening is obtained for GAL ($E(V-I)\sim 0.14$). The agreement with
the final adopted E(V-I) value for this Field (see below) is only marginal and
this reddening estimate has to be regarded as the most uncertain of the whole
set presented in table 2.

\subsubsection{Matching sequences and HBs}  

A more reliable way to estimate the differential reddening between SGR34 and
the others SDGS Fields can be achieved by the comparison of the mean position
and blue edge in color of the vertical blue sequence of the contaminating
population described in sec. 3 (hereafter VBS).
This method has been already discussed and applied to Sgr by IGWIS. Here we
slightly refine the procedure trying to simultaneously match the VBS
distribution with a color shift $\Delta E(V-I)$ and the magnitude of the peak 
of the Sgr HB population with a corresponding magnitude shift 
$\Delta A_V=1.93 \times \Delta E(V-I)$. Finding a $\Delta E(V-I)$ value that
satisfactorily matches both the constraints provides also support to the
statement that differential reddening is at the origin of the apparent shift
eventually detected. 

In figure 9, the results of the above procedure are reported. In left panels
the color distribution of the blue sequence stars (selected to have 
$16.5<V<19.5$ and $0.6<V-I<0.95$) are displayed. Dashed line histogram
refer to our ``reddening reference Field'', i.e. SGR34, whilst solid line
histograms refer to the SGR12 (upper panel), SGRWEST (middle panel) and GAL
(lower panel) samples, respectively,  after the appropriate $\Delta E(V-I)$ 
shift has been applied in order to match the compared distributions. 
The same symbols are
adopted for right hand panels, presenting the histograms of the stars in the
Sgr HB region (selected to have $17<V<19$ and $0.95<V-I<1.2$). The shifts have
been computed in the following way, taking, as an example, the comparison
between SGR34 and SGR12:

$$ \Delta E(V-I) = E(V-I)_{SGR34} - E(V-I)_{SGR12} $$

$$ \Delta A_V = {A_V}_{SGR34} - {A_V}_{SGR12} $$

In the cases of SGR12 and SGRWEST the match is excellent, and it can be 
concluded that the tiny shifts observed are probably due to slight reddening 
differences.
Though the reddening differences found (reported in the left panels of fig. 9)
are, at most, only marginally significant, we decided to adopt them since they
provide such a good match. This test also shows that any
distance difference between different regions along the major axis due to the
possible inclination of Sgr with respect to the line of sight is negligible,
over the scale sampled by SDGS ($4^o$).
    
\begin{figure*}
 \vspace{20pt}
\epsffile{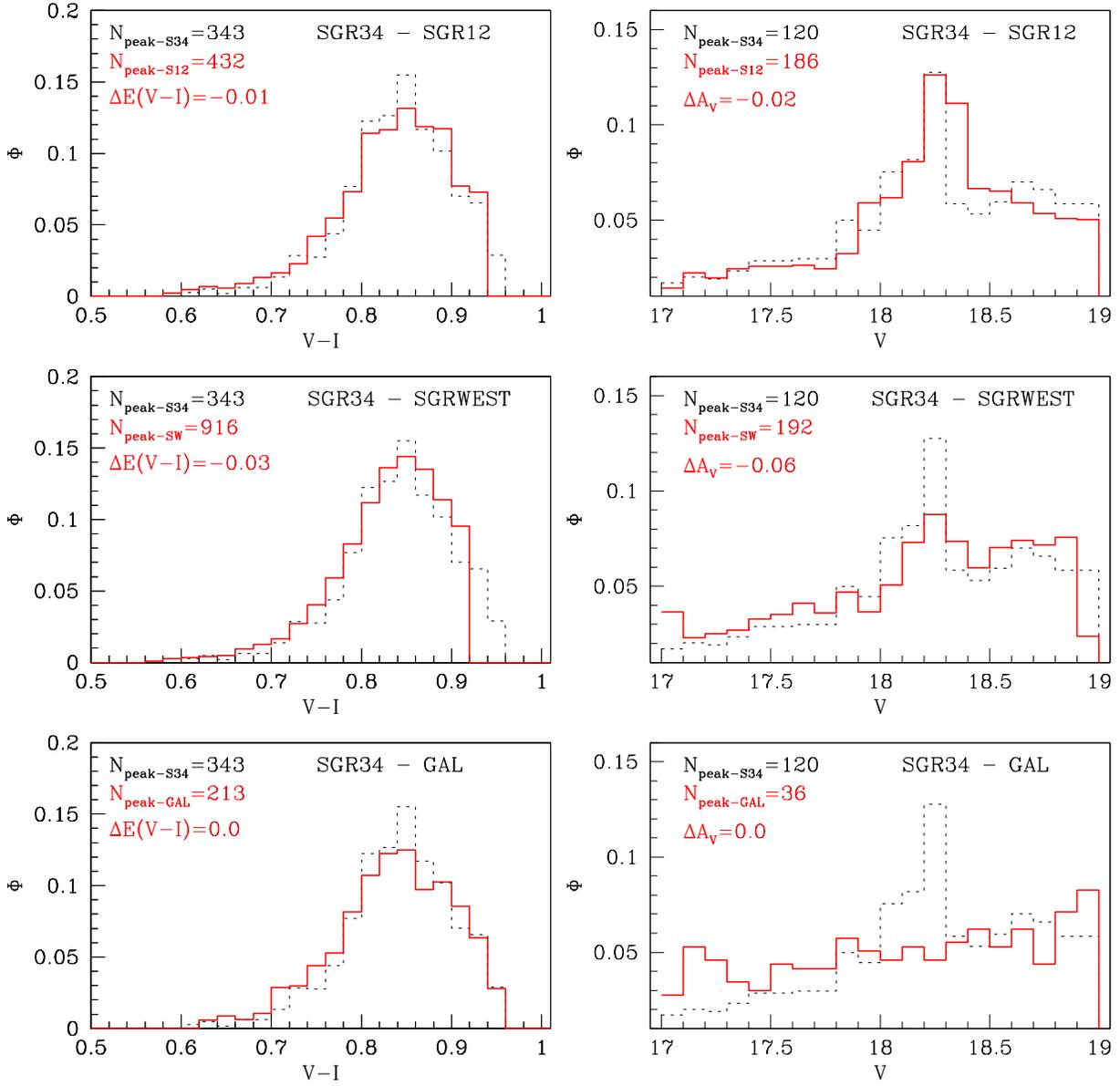}
 \caption{Differential reddening between SDGS Fields. Histograms in color (left
 panels) and in V magnitude (right panels) are presented for the Vertical Blue
 Sequence (VBS) of the contaminating population and the Sgr HB, respectively. 
 All histograms are normalized to the 
 total number of considered stars. In each panel the histograms obtained in
 the considered Field (solid line) has been shifted to match the corresponding
 distribution obtained for the SGR34 Field (dashed line). The applied shifts
 ($\Delta E(V-I)$ and $\Delta A_V=1.93 \times \Delta E(V-I)$)in color and
 magnitude are reported in each panel, together with the number of stars in the
 most populated bin ($N_{peak}$).}  
\end{figure*} 

In the case of GAL, the independent check provided by the match between HB 
peaks is obviously impossible and we are forced to consider only the color
distributions of VBSs. We finally adopt 
$\Delta E(V-I)_{GAL}=0.0$, but this result bear a considerable uncertainty.

Table 2\footnote{Table 2 specifications:
 
$^{\ast}$ The errors quoted for the SDGS Fields are the
standard deviation around the average E(V-I) values for the sub-fields. On the
other hand, the errors reported for the E(V-I) estimates for Sgr GCs are
derived assuming a relative uncertainty of $16 \%$, typical of ASRM.

$^{\star}$ We added to the error assumed from MUSKKK an arbitrary amount of 0.01
mag for the three Sgr Fields and of 0.02 for the GAL Field to account for the
uncertainties in the procedure used to derive differential reddenings.

$^1$ From MUSKKK. $^2$ Adopted, from MUSKKK. $^3$ From SL95. $^4$ From MoAL.
$^5$ From Sarajedini \& Layden (1997).}
presents a synoptic view of the reddening estimates for the SDGS Fields 
(from ASRM, from previous direct measures and the ones obtained above) 
and for the fields
containing Sgr globulars (from literature): Ter 8 ($l^o=5.76 ; b^o= -24.56$), 
Ter 7 ($l^o=3.39 ; b^o=-20.07$), M54 ($l^o=5.61 ; b^o=-14.09 $) and 
Arp 2 ($l^o=8.55 ; b^o=-23.27 $). 
All estimates are consistent with no (or very little) reddening differences 
across the region of the \sgr sampled by SDGS. 
Interstellar extinction appear rather homogeneous also on the wider scale
covered when directions toward Sgr GCs are also considered 
($\sim 10 ~deg$ across). 
  
\begin{table}
 \centering
 \begin{minipage}{140mm}
  \caption{Reddening estimates for SDGS Fields and Sgr GCs.}
  \begin{tabular}{@{}lccc@{}}
Field         &                           &    E(V-I)     &                  \\
              &     ASRM$^{\ast}$      &   Direct Estimates   & SDGS$^{\star}$\\
&&&\\
SGR34         & $0.20 \pm 0.01$ & $0.22 \pm 0.04^1$ & $0.22 \pm 0.04^2$ \\
SGR12         & $0.24 \pm 0.01$ & $0.21 \pm 0.03^3$ & $0.23 \pm 0.05$   \\
SGRWEST       & $0.22 \pm 0.01$ &        ---        & $0.25 \pm 0.05$   \\
GAL           & $0.14 \pm 0.02$ &        ---        & $0.22 \pm 0.06$   \\
&&&\\
M54           & $0.24 \pm 0.04$ & $0.21 \pm 0.03^3$ &      ---         \\
Ter 8         & $0.22 \pm 0.04$ & $0.19 \pm 0.05^4$ &      ---         \\
Ter 7         & $0.14 \pm 0.02$ & $0.11 \pm 0.05^5$ &      ---          \\
Arp 2         & $0.17 \pm 0.03$ & $0.16 \pm 0.03^5$ &      ---          \\
\end{tabular}
\end{minipage}
\end{table}

\section{Population differences between the three Fields}

While detailed studies of the age and metallicity of Sgr populations needs
accurate decontamination of the samples from foreground stars, many useful
statements can be made on the CMDs presented in fig. 5, 6, and 7. In particular
we used the wide spatial coverage of SDGS to search for possible differences
in the Sgr stellar content in different regions of the galaxy.

\begin{figure*}
 \vspace{20pt}
\epsffile{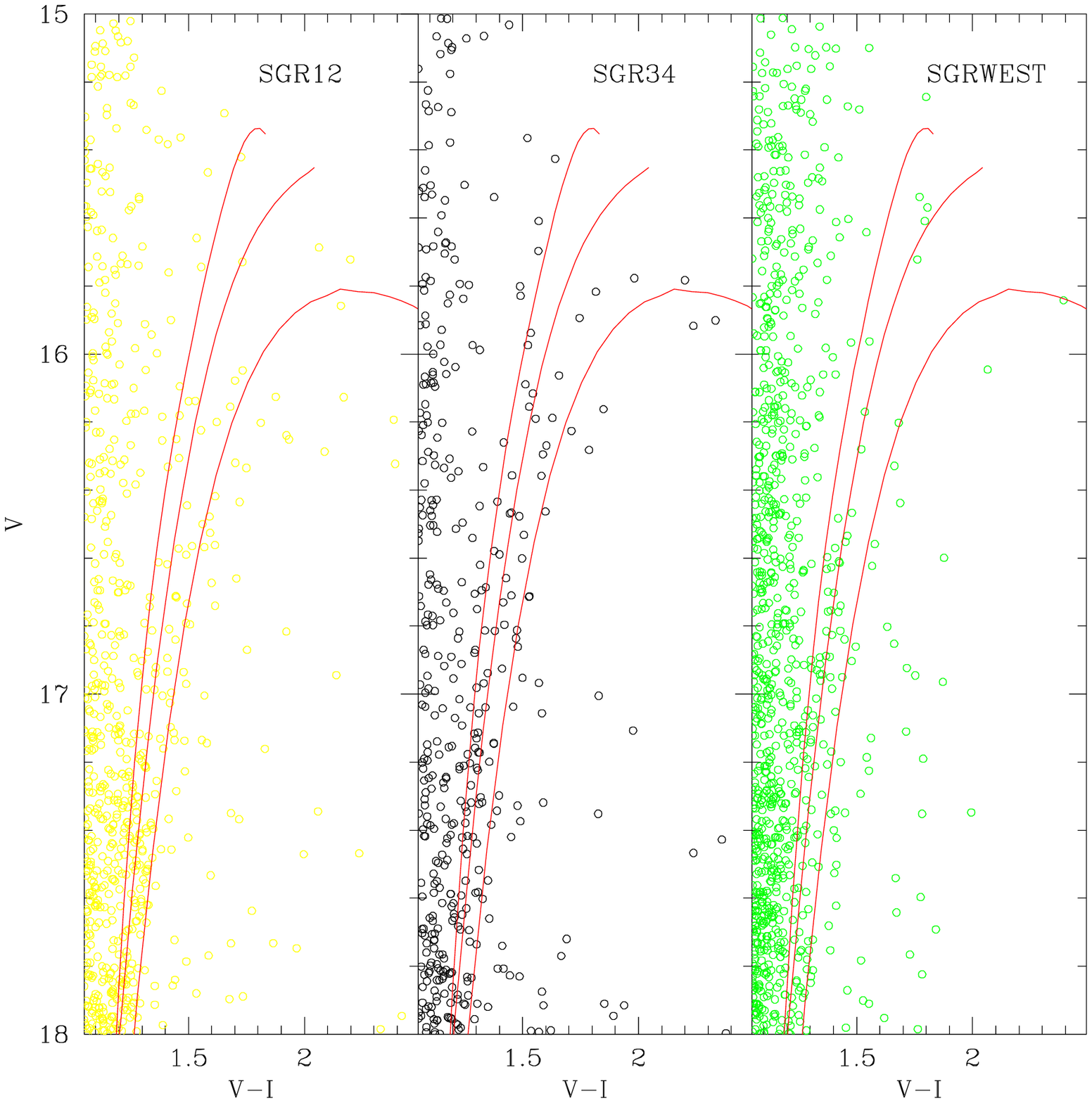}
 \caption{The giant branch of the Sgr Pop A in the three SDGS Fields. SGR12 and
 SGRWEST stars were corrected for differential reddening with respect to SGR34,
 according to the values found in sect. 4.1.1. The solid lines are the mean RGB
 loci for three galactic globular cluster (from right to left): 47 Tuc
 ($[Fe/H]=-0.71$), NGC1851 ($[Fe/H]=-1.29$) and NGC 6752 ($[Fe/H]=-1.54$), from
 DCA1990. The ridge lines have been reported to the same distance of Sgr
 (assuming $V_{HB}(Sgr)=18.22$) and reddened in order to match the reddening of
 SGR34.}
\end{figure*} 

\subsection{Constraints on metallicity from the RGB location}

Fig. 10 shows a zoomed view of the CMD containing the brightest part of
Sgr RGB in the three SDGS Fields (SGR12, left panel; SGR34, central panel;
SGRWEST, right panel). The small differential reddening corrections described
above have been applied to SGR12 and SGRWEST samples to report them to the same
reddening as SGR34. For $V<17$ and $V-I>1.3$ field contamination is not a 
major concern, Sgr RGB clearly stand out, being one of the key feature
which led to the discovery of the Sgr galaxy (see IGI-I and IGI-II). 

The solid lines in fig. 10 are RGB ridge lines of (from right to left) 47 Tuc
($[Fe/H]=-0.71$), NGC1851 ($[Fe/H]=-1.29$) and NGC 6752 ($[Fe/H]=-1.54$), from
Da Costa \& Armandroff 1990, shifted at the Sgr distance and reddening. 

The most remarkable feature of fig. 10 is the wide color spread of the brightest 
part of the RGBs, a feature commonly interpreted as indication of significant
abundance spread among stars (MUSKKK, SL95, MAL). Any possible spread induced
by photometric stochastic errors (few hundredths of mag) should have a 
negligible effect since it is much smaller than the observed color spread.

The superimposed ridge lines (the same used also by MUSKKK - a further confirm 
of the excellent agreement between SDGS and MUSKKK photometries) seems to
nicely bracket the observed distributions and it can be concluded that the
metallicity spread is of order $0.7 - 0.8 ~dex$. Such spread is rather large 
with respect to what found in other dSph galaxies [Ursa Minor 
(Olzewski \& Aaronson 1985), Draco (Grillmair et al. 1998), 
Carina (Smecker-Hane et al. 1994)].

In the left panel of fig. 10 (SGR12) a few stars around 
($V\sim 16.5; ~V-I\sim 2$) suggest the possible presence of an additional RGB
sequence significantly more metal rich than 47Tuc, which has no counterpart in
the other panels. Such an evidence, if confirmed, would 
support the claims (based on rather marginal clues) by SL95 and
MAL for the possible presence of a slight metallicity gradient toward the
center of density of Sgr. Further evidences in favor of this hypothesis will 
be given below.

\subsection{The Horizontal branch}

As previously noted, a predominantly red HB around $V\sim 18$ and $V-I \sim 1$
can be easily identified in the CMD in fig. 5, 6, 7. Moreover the signature
of this features clearly emerges from star counts over the foreground
contamination (see fig. 9). In this section we will determine the mean level 
and the spread (in V magnitude) of the HB.   

The HB
dispersion is (potentially) an important observable in this case since it is 
the result of 
many factors (photometric errors, evolutionary effects, mix of HB stars of 
different metallicities) the main one being perhaps the extension of the Sgr 
galaxy along the line of sight (see IGWIS). Furthermore, also major features 
in the HB color distribution can probably be depicted out of the confusion
produced by foreground star contamination. The left panels of fig. 11 show the
V histogram of the box ($17.4 < V < 18.6; ~0.95<V-I<1.15$) which presumably
contain the region of the Sgr HB less affected by contamination (i.e. the red
part). RR Lyrae stars, probably present in our sample but with V and V-I
measured at random phase, lie in a region of the CMD where many foreground
stars are also found. 
Comparison of
the upper three panel of fig. 11 with the lower one (GAL, corrected for the 
factor accounting for the smaller sampled area) clearly states that Sgr 
population is largely dominant in this box.
An evident bell shaped distribution is present between
$V\sim 18$ and $V\sim 18.5$: we compute the mean V magnitude ($<V_{HB}>$) and 
standard deviation ($\sigma_{V_{HB}}$) over this
range for the SGR12, SGR34, and SGRWEST samples. 
The results are reported in table 3, together with $3\sigma$ errors 
on the mean
($3\sigma_{V_{HB}}/\sqrt{N_{star}}$) that we take as conservative error on the 
$<V_{HB}>$ estimates. 
All $<V_{HB}>$ and $\sigma_{V_{HB}}$ estimates are consistent within the errors.
In the last row of table 3 are reported the average values over the three 
Fields, calculated after the application of the differential reddening
correction to SGR12 and SGRWEST with respect to SGR34. 
The mean value is  $<V_{HB}>=18.22\pm 0.02$, consistent
within errors with all previous estimates (MUSKKK, MAL, LS97). A particularly
good agreement is achieved with SL95 ($<V_{RHB}>=18.25\pm 0.05$, for the red HB)
and IGWIS ($<V_{HB}>=18.25\pm 0.05$, from their fig. 2). 
The $<V>_{RR Lyrae}=18.16\pm 0.05$
estimate obtained by MUSKKK from a small sample of RR Lyrae variables toward
SGR34 seems marginally inconsistent with our estimate, above all if it
is taken into account the fact that RR Lyrae are expected to be slightly fainter
than very red HB stars (see DCA90, for example) at the same metallicity.
However, two out of seven of the MUSKKK RR Lyrae are much brighter than the
mean HB level (see their fig. 2) and have probably a great weight on the final
average $<V>_{RR Lyrae}$ value. A deeper analysis of the RR Lyrae sample of
MUSKKK is performed by MKSKKU. The average of the {\em bona fide} RR Lyrae
from table 8 of MKSKKU (i.e. SV1, SV2, SV3, SV4, SV6 and SV7) gives 
$<V>_{RR Lyrae}=18.26 \pm 0.05$, in agreement with our result.
Furthermore it must be considered that red HB stars and RR Lyrae in Sgr almost
certainly have NOT the same metallicity, so comparisons are significant only in
a rough sense. Finally, Mateo et al. (1996) found $18.22 \le V_{RR Lyrae} \le
18.36$ for 3 RR Lyrae associated to Sgr in the line of sight toward M55. 

At present it is not possible to disentangle the different factors contributing
to the magnitude dispersion of the HB stars. The only firm conclusion that can
be drawn at the moment is that $\sigma_{V_{HB}}$ is clearly larger than 
photometric error at that magnitude.    

\begin{table}
 \centering
 \begin{minipage}{140mm}
  \caption{HB levels and dispersion for SDGS Fields.}
  \begin{tabular}{@{}lcc@{}}
Field         &  $<V_{HB}> ~\pm 
3\sigma_{V_{HB}}/\sqrt{N_{star}}$&$\sigma_{V_{HB}}$ \\
&&\\
SGR34         & $18.23\pm 0.02$ & $0.12$\\
SGR12         & $18.26\pm 0.02$ & $0.12$\\
SGRWEST       & $18.25\pm 0.02$ & $0.13$\\
&&\\
average$^{\dag}$ & $18.22\pm 0.02$ & $0.12 \pm 0.01$\\
\end{tabular}

\footnotesize{$^{\dag}$ After corrections for differential reddening.}

\end{minipage}
\end{table}

\subsubsection{Distance Modulus}

A straightforward way to determine Sgr distance from $<V_{HB}>$ is to find the
differential distance modulus with respect to a well studied globular cluster
of comparable metal content and HB morphology. The ideal subject in this case
seems to be 47 Tuc, the prototypical metal rich globular. First of all we
correct for interstellar extinction the average $<V_{HB}>$ value, 
obtaining $<V_{HB}>_0=17.80 \pm 0.05$ for Sgr. Adopting for 47 Tuc,
$<V_{HB}>_0=14.06$ and $(m-M)_0=13.51$ from DCA90 we obtain:
 
$$\Delta(V_{HB})=V_{HB}(Sgr) - V_{HB}(47 Tuc) = 3.74$$

and finally:

$$<(m-M)_{Sgr}>_0=17.25 ^{+0.1} _{-0.2}$$

corresponding to $\sim 28 ^{+2} _{-4} Kpc$

The asymmetrical error bar is due to the correction made by DCA to account for
the expected higher luminosity of 47 Tuc red HB stars with respect to RR Lyrae,
on which their distance scale is based. They assume that HB stars of 47 Tuc are
0.15 mag brighter than RR Lyrae. While correction of this amount are commonly
used the uncertainty on the actual value is comparatively large, since
correction of $\sim 0.05$ mag have also been suggested. 
Taking into account the large uncertainty in this correction (that was never
considered before in deriving the distance to the \sgr; see IGWIS, MUSKKK, 
MKSKKU) it can be useful to provide also an estimate of distance modulus 
{\em without} the DCA90 correction, i.e.: 
$(m-M)=17.1 \pm 0.2$($\sim 26 \pm 3 Kpc$), in good agreement with published 
measures (IGWIS, MUSKKK, MKSKKU). However, it has to be stressed that some
correction is necessary to report the mean level of the red HB to the mean
level of RR Lyrae as is apparent from fig. 6 of MKSKKU. The marginal
disagreement between our estimate (corrected for the quoted effect) 
$(m-M)_0=17.25 ^{+0.1} _{-0.2}$ and the distance modulus derived by MKSKKU 
$(m-M)_0=17.02 \pm 0.19$ is due do the different extinction law assumed.
In fact, from the same $E(V-I)=0.22 \pm 0.04$ they derive $A_V=0.55$ while in
the present work $A_V=0.42$ is assumed, according to RL. 

Finally we recall
that while the error on the estimate by MKSKKU take into account also the
uncertainty on the slope of the $M_V(RR Lyrae) ~vs. ~[Fe/H]$ relation, our
error bar do not include this effect.

\begin{figure*}
 \vspace{20pt}
\epsffile{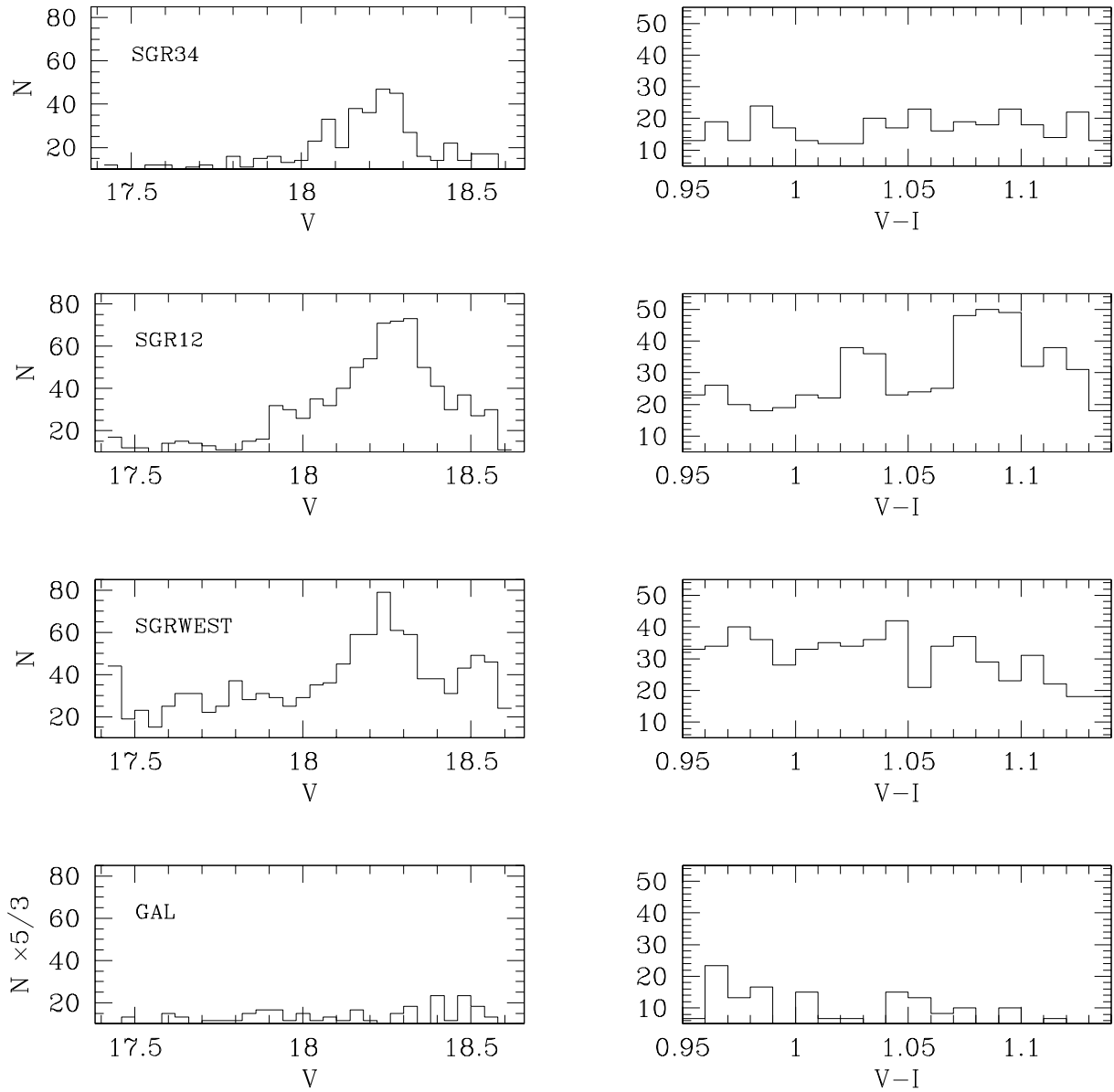}
 \caption{Histograms in magnitude (left panels) and color (right panels) for 
 the stars in the HB box (see text). 
 Comparing the two upper-right panels (SGR34 and SGR12, respectively) 
 the redder morphology of the HB in SGR12 is
 evident: a clear bump is present at $V-I\sim 1.07 - 1.10$ which has no
 counterpart in the SGR34 and SGRWEST histograms.}  
\end{figure*} 

\subsection{HB morphology}

The right panels of fig. 11 report the color distribution of stars in the HB box
described above. While HB stars distribution seems very flat in
this range of colors for SGR34 and SGRWEST,
the histogram of the SGR12 sample presents a clear bump at 
$V-I\sim 1.07 - 1.10$ having no counterpart in the SGR34 and SGRWEST 
histograms. SL95 argued for a ``double'' HB morphology of Sgr stars in this
region: a Red Clump HB, that they associate with a population 10 Gyr old and 
with $[Fe/H]\sim -0.5$, plus an ``ordinary '' red HB associated with an older
and more metal poor ($[Fe/H]\sim -1.3$) population, similar to globular
clusters like NGC 362, Pal 4 and Eridanus.

SGR12 color histogram of fig. 11
seems to support this view. In particular, the presence of a second peak at 
$V-I\sim 1.02 - 1.04$ may suggest a true bimodality of the HB morphology.
However the significance of the apparent bimodality cannot be
firmly established at present, and has to be regarded as a marginal evidence, at
least unless accurate decontamination have been carried on (Paper II).

On the other hand the detected difference in HB morphology between SGR12 
and SGR34 and SGRWEST is very likely to be real. 
A Kolmogorov-Smirnov test states
that {\em the probability that SGR12 and SGR34 HB samples are drawn from the 
same parent population in color is less than $3 \%$}. 
Note that the presence of the
nearly uniform contaminating population in this region of the CMD tend to smear
out real differences in the distributions of Sgr stars, so the above figure can
be regarded as a strong upper limit for likelyhood of the quoted 
{\em null hypothesis}.
This results are unaffected if the SGR12R sample, instead of the whole
SGR12 one, is used.

The redder HB morphology coupled with the redder RGB discussed in sec. 5.1
suggest the presence of a metallicity (and age?) gradient from the outer
regions toward the center of density of the \sgr (SGR12).
 
\subsection{Number counts comparisons}

In the previous sections we presented CMDs based on large samples of stars
measured in three different region of the Sgr galaxy. The samples are populous
enough to be representative of the stellar population in each region of the
galaxy. In order to put in evidence (quantitatively) possible spatial
differences in stellar content among the surveyed Fields we compared star
counts in selected boxes in the CMDs. The boxes have been defined to select
regions of the CMDs {\em dominated by one clear-cut feature}, and so they are
expected to track unambiguously only {\em one} of the stellar components
identified on the SDGS CMDs (i.e. Sgr Pop A, Sgr Pop B or Milky Way
Bulge+disc). 

We identified five key features on the CMD. Three of them (the   
Sgr RGB, the red HB around $V\sim 18$ and the upper Main Sequence 
around $V\sim 21$ and $V-I\sim 0.75$, just above the TO point) 
have been taken as tracers of Sgr Pop A, since are clearly associated with 
this older population. The fourth selected feature is the Blue Plume described 
in sec. 3 and associated with Sgr Pop B, and the fifth is a portion of the 
vertical Bulge+disc sequence, tracing the degree of foreground contamination 
of the considered Field.
      
\begin{figure*}
 \vspace{20pt}
 \caption{The boxes selected for star counts of tracer populations are
 represented and labeled, superimposed on the CMD of an arbitrary subsample of
 SGR34 stars. Only a relatively low number of stars have been plotted to avoid
 boxes being hidden by ink blobs.}  
\end{figure*}

The boxes have been defined in order to avoid reciprocal contamination 
between the different CMD features. .
The boxes, labeled after the dominant population sampled, are clearly 
illustrated in fig. 12 and are defined as follows:

\begin{itemize}

\item RGB: $16.0 \le V \le 17.8$ and $1.25 \le V-I \le 1.7$ containing most of 
the Sgr Pop A Red Giant Branch.

\item HB: $18.0 \le V \le 18.5$ and $0.95 \le V-I \le 1.2$ containing most of 
the
Sgr Pop A Horizontal Branch, at least the part less affected by the
contaminating sequence.    

\item TO: $21.0 \le V \le 21.5$ and $0.65 \le V-I \le 0.8$ around the brightest
part of Sgr Pop A Main Sequence, just above the Turn Off point. 
In this case we can take advantage of the high
number of stars in this region of the CMD to minimize the dimension of the box
without any loss in statistical significance. So, the box edges are chosen with
particular care to avoid contamination from the large number of
foreground/background sources present
in the CMD for $V-I > 0.9$ at this magnitude level. The significance of star
counts in this box is nevertheless seriously weakened by two factors: {\em (a)}
the completeness is less than $50 \%$ for $V\sim > 21.$ in each of the SDGS
Fields, so the completeness correction is very large and uncertain; {\em (b)}
the completeness factor is not well defined at that magnitudes for the SGR12
and GAL samples, the faint edge of the box being very near to the detection
threshold of the photometry. Star counts in the TO box for these Fields must be
considered only as lower limits. However SGR34 and SGRWEST have very similar
completeness and limiting magnitude, so number counts comparisons between these
latter samples is probably significant and very interesting.

\item BP: $18.8 \le V \le 21.$ and $0.2 \le V-I \le 0.6$ containing
most of the Blue Plume sequence (Pop B). The faint edge of the box is close
the $C_f = 0.5$ limit, so some underestimation in star counts for SGR12 and GAL
(see above) can be expected.

\item BMS: (Bulge Main Sequence) 
$18.6 \le V \le 19.5$ and $0.75 \le V-I \le 0.95$.
The box select the foreground star sequence in a zone of the CMD where the
completeness is fair and the expected contamination by Sgr stars is virtually
null.

\end{itemize} 

\begin{table}
 \centering
 \begin{minipage}{140mm}
  \caption{Star counts in selected CMD boxes.}
  \begin{tabular}{@{}lccccc@{}}
Field         &  $N_{RGB}$ & $N_{HB}$ & $N_{TO}$ & $N_{BP}$ & $N_{BMS}$\\
&&&&&\\
GAL           &  34        &  200     &   283*   &  154	    &  1357  \\
SGR34         & 105        &  430     &  2940    &  306	    &  1235  \\
SGR12         & 160        &  756     &  2670*   &  648	    &  2183  \\
SGRWEST       & 140        &  753     &  1784    &  457	    &  4940  \\        
&&&&&\\
\end{tabular}

{\footnotesize * asterisks indicate uncertain values (see text)} 
\end{minipage}
\end{table}

For the star counts in the RGB and HB boxes the whole SGR12 sample was used,
while for the boxes below $V=18.5$ (TO, BP and BMS) the SGR12R sample 
was adopted, in order to prevent misleading effects from the modest quality of
the photometry at faint magnitudes of subfields SGR12D and E (see sec. 2).
Star counts from the SGR12R and GAL sample have been normalized to the area of 
a whole strip ($\simeq 9 \times 35 ~arcmin^2$), multiplying the observed 
number counts by a 5/3 factor.
All the counts were corrected for completeness according to the relations shown
in fig. 3.

The corrected number counts in the different boxes are reported in table 4 and 
displayed in fig. 13 for GAL and for the Sgr Fields ordered with increasing 
galactic latitude 
(labels with the approximate latitude of the Sgr SDGS fields are also reported). 
The error bars have been obtained applying error
propagation, adding Poisson 1-$\sigma$ uncertainties in number counts 
to an assumed conservative error of $10 \%$ in the applied completeness 
corrections.     

The upper four panels of fig. 13 should be interpreted with some caution, since,
as we will show below, the contamination of star counts by foreground stars can
be very different from Field to Field. However some general features are readily
evident:

\begin{enumerate} 

\item Star counts in boxes containing tracers of Sgr population 
(RGB, HB, BP and TO - upper four panels) are significantly higher in 
Sgr Fields with respect to GAL.

\item The trend of star counts with galactic latitude is
remarkably similar for the upper three panels of fig. 13, the first two 
reporting tracers of Sgr Pop A and the third reporting {\em a Pop B tracer}.

\item Star counts in the TO box are
significantly higher in SGR34 than in SGRWEST. The dashed square around the
point representing TO star counts in SGR12 has the aim of emphasizing the
large uncertainty affecting this estimate (see above). The {\em true}
$N_{TO}(SGR12)$ could be significantly higher than what reported. 

\end{enumerate}

\begin{figure*}
 \vspace{20pt}
\epsffile{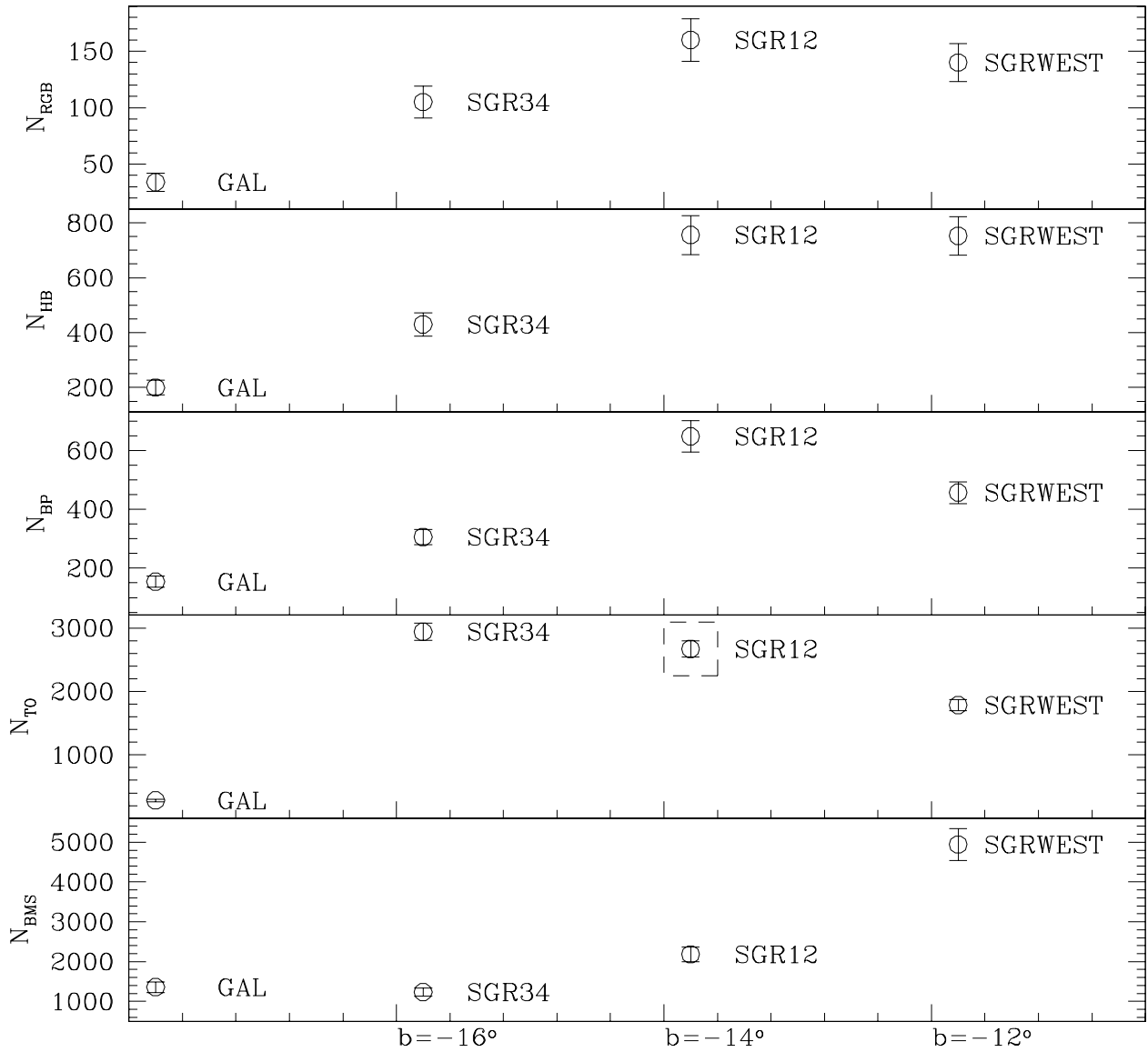}
 \caption{Number of stars counted in the selected boxes (see fig. 12) for each
 SDGS Field plotted as a function of galactic latitude. The dashed square
 around $N_{TO}$ for the SGR12 Field indicate that this number should be
 considered as a lower limit.}  
\end{figure*} 

However, the most important result of fig. 13 resides in the lowest panel,
reporting star counts in the CMD box tracing the sole contaminating population.
The very steep gradient in the number of foreground stars between $b=-16^o$ and
$b=-12^o$ can be quantitatively appreciated: this number nearly doubles 
from the position of SGR34 to that of SGR12, and it roughly doubles again from
SGR12 to SGRWEST. Furthermore it result evident that the GAL Field represents
optimally the contaminating population for SGR34 while it significantly
underestimates contamination for SGR12 and SGRWEST. This latter fact must be
taken into account also when CMD statistical decontamination is attempted (SDGS
Paper II). 

\subsection{Decontaminating star counts}

Ideally, one can effectively remove much of the contamination in star counts
by subtracting the star counts in a given box of a CMD containing only stars
belonging to the Contaminating Population (CP) to the star counts in the 
same box of a CMD containing Sgr+(CP) stars, 
once a sample with the following characteristics is available:

\begin{enumerate}

\item composed solely of CP stars

\item representative of the CP present in the CMD containing both Sgr and CP
stars (i.e., same CP star density in each box).

\end{enumerate}

In order to quantify how much the GAL sample is representative of the CP in the
different SDGS Fields we defined the ratio
 
$$\phi_{Sgr Field}=N_{BMS}(Sgr Field)/N_{BMS}(GAL)~~~~.$$

This index approximatively represents the degree of contamination in each Field
with respect to the GAL field. It is obtained: $\phi_{SGR34}= 0.9$, 
$\phi_{SGR12}=1.6$ and $\phi_{SGRWEST}= 3.6$. 

So, it can be concluded that the GAL sample seems fairly representative of the 
CP in SGR34 (see also fig. 13, lower panel).

\begin{figure*}
 \vspace{20pt}
\epsffile{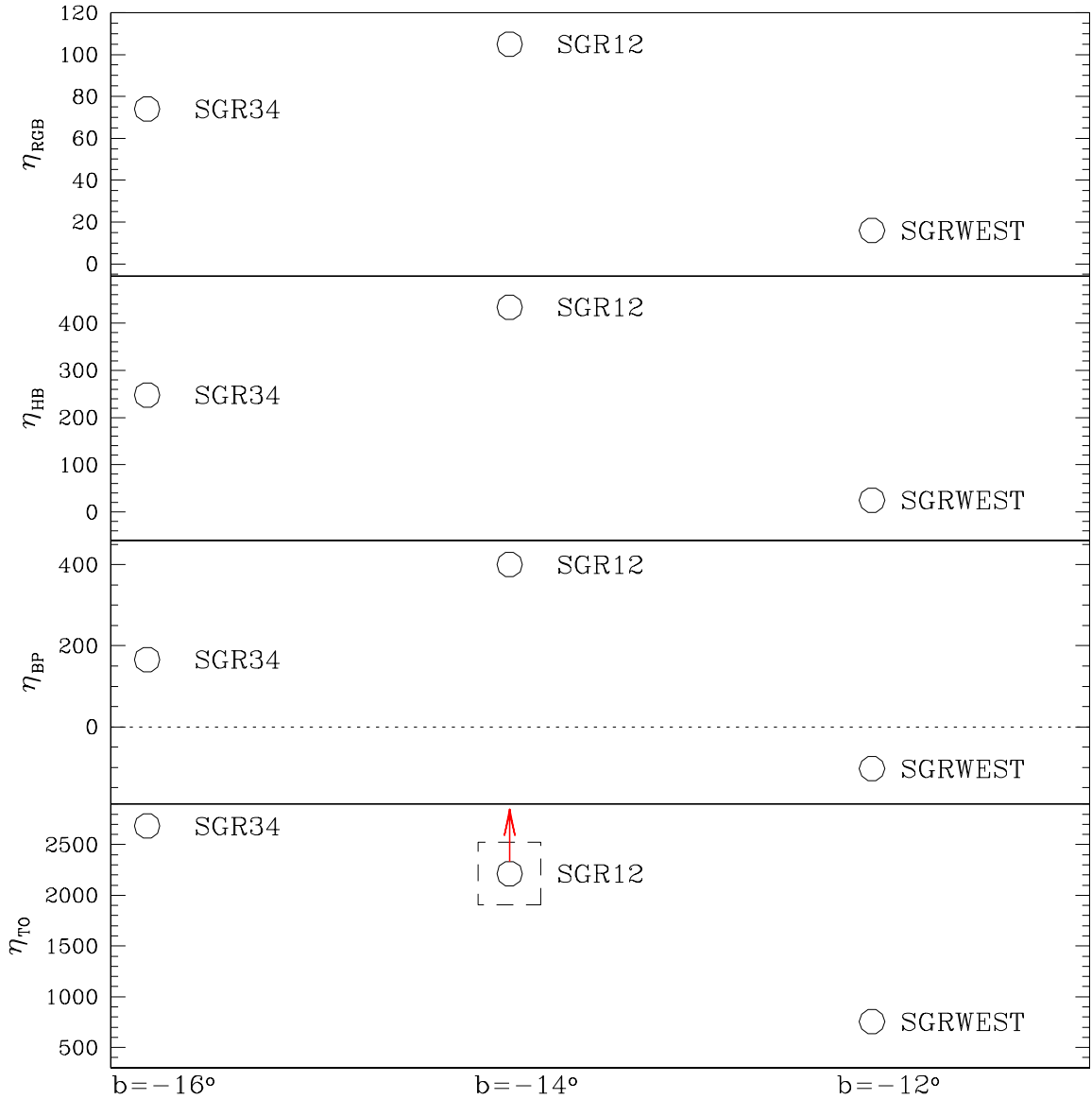}
 \caption{The same as figure 13, for decontaminated star counts (see text).
 The arrow indicates that the reported $\eta_{TO}(SGR12)$ is a lower limit.}  
\end{figure*} 

Furthermore, we can use the $\phi$ ratio to estimate the expected number of CP 
star in a given box for the SGR12 and SGRWEST samples.

By multiplying
the number of star in a given box in the GAL sample by the appropriate
$\phi$ ratio we can obtain such estimate, and then subtracting it 
to the counts in the same box of the CMD of a Sgr Field we get the number of
Sgr stars in that box for the given Sgr Field. Let's call this number
$\eta_{box}(Sgr Field)$, defined as follows (taking as an example the HB
box and the SGR34 Field):

$$\eta_{HB}(SGR34) = N_{HB}(SGR34)-[N_{HB}(GAL)\times \phi_{SGR34}]~~~.$$ 

where the term included in the square brackets is an estimate of the number of 
CP stars present in the HB box of the SGR34 CMD and $\eta_{HB}(SGR34)$ is
in principle, the number of {\em true} Sgr stars in the HB box of the
SGR34 CMD.

We can reasonably expect that passing from $N_{box}$ to $\eta_{box}$ much 
of the contamination is removed. Given the strong gradient of CP star density
in the region of sky occupied by \sgr, the adopted approach is the most viable
to deal with Sgr star counts. The only alternatives coming to mind are:

\begin{enumerate}

\item to estimate the degree of local contamination by adopting a Galaxy model
[as, for example, Bachall \& Soneira (1980, 1984)]. However this approach is
prone to large uncertainties when applied to such small spatial scales.

\item To acquire a large number of control fields in different positions (for
example, a sequence of fields along a line parallel to Sgr major axis, some
degree apart) such to have a significant
probability to get representative CP samples for each of the Sgr fields to
analyze. This strategy is very expensive in terms of observation time and still
the ``degree of representativity'' must be somehow estimated ``a posteriori''.

\end{enumerate}

As correctly pointed out by the referee, the main pitfall of the adopted
approach resides in the implicit assumption that the latitude density gradient 
of
bulge and disc stars is similar (in the considered range, i.e. from $b=-16^o$
to $b=-12^o$) or, in other words, that the CMDs of the CP stars in the
different fields differ only in the number of stars. This is indeed a rather
crude approximation that could introduce significant  systematic errors in 
the ``decontaminated'' number counts. This possible occurrence has to be taken
into account while interpreting the results presented in the following
subsections (sec. 5.1.1 and 5.1.2).   

The caveats described in the previous section concerning $N_{TO}$ 
obviously apply also to $\eta_{TO}$, in particular for the SGR12 sample.

The plots $\eta_{box} ~vs. ~ Sgr ~Field$ are displayed in fig. 14.
Labels and symbols are the same as fig. 13, 
obviously without the GAL points since, by definition, $\eta_{box}(GAL)=0$ 
for any box. 
No error bar is reported since the indeterminacy in the procedure from $N$ to 
$\eta$ is tied only to the validity of the assumption of ``representativity'',
so intrinsically hard to quantify.  

\subsubsection{Asymmetry in density distribution}

The upper two panels of fig. 14 show the behaviour of star count for the most
reliable tracers of Sgr Pop A, i.e. $\eta_{RGB}$ and $\eta_{HB}$. 
The trend is
very similar in both panels: there is an enhancement in Sgr stellar density
from SGR34 to SGR12 by a factor $\sim 1.6$ while the enhancement between
SGRWEST and SGR12 is of nearly one order of magnitude. 
There are two possible explanation
for this fact: (1) the tidal limit of the \sgr occurs at lower galactic
latitudes with respect the SGRWEST Field (i.e. the field is devoid of Sgr
stars), or (2) there is a clear asymmetry in
the distribution of Sgr stars along the major axis, the region nearer to the
Galactic Bulge suffering strong star depletion (possibly related to ongoing
disruption by tidal interaction with the Galaxy). The first
hypothesis is ruled out by the lower panel of fig. 14:
$\eta_{TO}(SGRWEST)=754$, so there is a clear excess of stars in the TO
box of SGRWEST with respect to GAL. Then, it must be concluded that Sgr stars
are still significantly present at $b=-12^o$ (as already stated by IGWIS) but
the Sgr star density is probably much lower with respect to the symmetrical 
SGR34 Field 
[see also the comparison between $\eta_{TO}(SGRWEST)$ and $\eta_{TO}(SGR34)$]
and there is no symmetry in star distribution with respect to the center of 
density ($\sim$ SGR12), at least along the major axis. 

The decreasing in star counts from the SGR12 Field to the SGR34 Field is in 
rough accord with what expected for a King Model distribution 
(assuming $C\sim 0.5$, $r_t\sim 4^o$ and $r_c\sim 1.25^o$, as estimated by 
IGWIS) but the strong star deficiency in SGRWEST seems to rule out 
King Models as viable density distributions for Sgr.
The clumpy nature of the Sgr galaxy has already been noted 
[IGI-II, IGWIS, Mateo et al. (1996)] and the SGR34 Field is centered on one of
the main subclump (IGI-II). However the fall of star counts toward SGRWEST
is very steep and suggest a remarkable depletion of Sgr stars in this region.

We recall again that decontaminated star counts can be affected by unaccounted
variations in the CP composition from field to field. On the other hand, it has
to be considered that the same result is found from star counts in very
different regions of the CMD. One would expect that the various windows 
(TO,HB,RGB,BP) have to be affected in different ways if the distribution of CP
stars in the CMD changes significantly from $b=-16^o$ to $b=-12^o$. So, while
the actual extent of the asymmetry is quite uncertain, it seems unlikely
that differences in the relative abundance of bulge and disc stars could
produce such an effect in {\em all} the selected windows.

\subsubsection{The distribution of the Blue Plume population} 

If the Blue Plume stars are associated to a younger generation of stars (Pop B)
their distribution in the galaxy may be very different from the main
population. Presently active dwarf galaxies are observed to experience very
localized burst of star formation on scales of the order of few $\sim$ 100 pc 
(Tosi 1993, Hunter 1998), significantly smaller than the scale sampled by 
SDGS, ($\sim 2 ~Kpc$ from SGR34 to SGRWEST). 
The third panel (from up to down) of fig. 14 shows
that this is not the case for Sgr, the general behaviour of $\eta_{BP}$ 
remarkably resembling the trend of the Pop A tracers $\eta_{RGB}$ and 
$\eta_{HB}$, at least in the transition from SGR34 to SGR12. 

On the other hand
$\eta_{BP}(SGRWEST)=-103$, i.e. the contribution by the contaminating 
population in this box is expected to be higher than actually observed in the
SGRWEST Field. This result is ambiguous since can indicate that (a) the BP
population is absent in this region, or (b) the number of BP stars is too low to
survive the decontamination because of the low density of Sgr in this region. 
We put it in evidence since, in our view, it surely deserves further check
( a deeper photometric study or a survey of bright Pop B tracers, as Carbon
Stars).

Finally we can try to roughly estimate the contribution of Pop B stars to the
whole Sgr population. The most suitable sample for this purpose is SGR34.
The factor
$f={\eta_{BP}\over{{(\eta_{TO}+ \eta_{BP})}}}\simeq 0.06$ can be taken as an
upper limit for the relative contribution of BP stars in this Field.

\section{Is there a metal poor (and old) population in Sgr?}

The only direct evidence of the presence of a (relatively) old and metal poor
population in the \sgr comes from the detection of RR Lyraes (MUSKKK, Mateo et
al 1996, Alcock et al. 1996). As pointed out by IGWIS, the mere existence of 
such
stars is indicative of age $\sim > 10-12 ~Gyr$ and metallicity $[Fe/H] \sim <
-1.5$. However, at least two member of the Sgr globular cluster system are
significantly older and more metal deficient than the above figures, i.e. Ter 8
($[Fe/H]=-2.0$ see MoAL, Da Costa \& Armandroff 1995) and M54 (
$[Fe/H]=-1.79$, see MAL and SL97). So, if the formation of GCs is not
completely decoupled from the star formation in the host galaxy (i.e. Ter8 and
M54 formed many Gyrs before the onset of star formation in Sgr), 
some fraction of stars with
age and metallicity similar to the above quoted globulars should be present 
also in the Sgr field.
This hypothesis can be tested looking in the CMDs for features typical of old
and metal poor globular clusters.
In particular we used the mean loci defined by stars in the CMD of Ter 8 
 to test if a ``Ter 8-like'' population is present in the Sgr galaxy.
We adopted the ridge line by MoAL, since the instrumental set up of MoAL is
the same used for the SDGS. 

\begin{figure}
 \vspace{20pt}
 \caption{The Ter 8 ridge line (by MoAL) is superposed to the CMD of a
 subsample of stars in the SGR34 Field, after the corrections specified in the
 text.}
\end{figure}

The SDGS and MoAL CMDs were corrected for reddening and a shift $\Delta
V_0=+0.25$ were applied to the Ter 8 ridge line in order to roughly
match HB levels. This shift correspond to
reporting Ter 8 to the average distance of Sgr, since the cluster seems to lie
slightly in foreground along the line of sight with respect to the 
main body of the host galaxy (see IGWIS and Da Costa \& Armandroff 1995).
The correction takes into account also the expected difference between the
level of the very
blue HB (Ter 8) and the stubby red one of the \sgr (see sec. 5.2.1). 

\begin{figure*}
 \vspace{20pt}
\epsffile{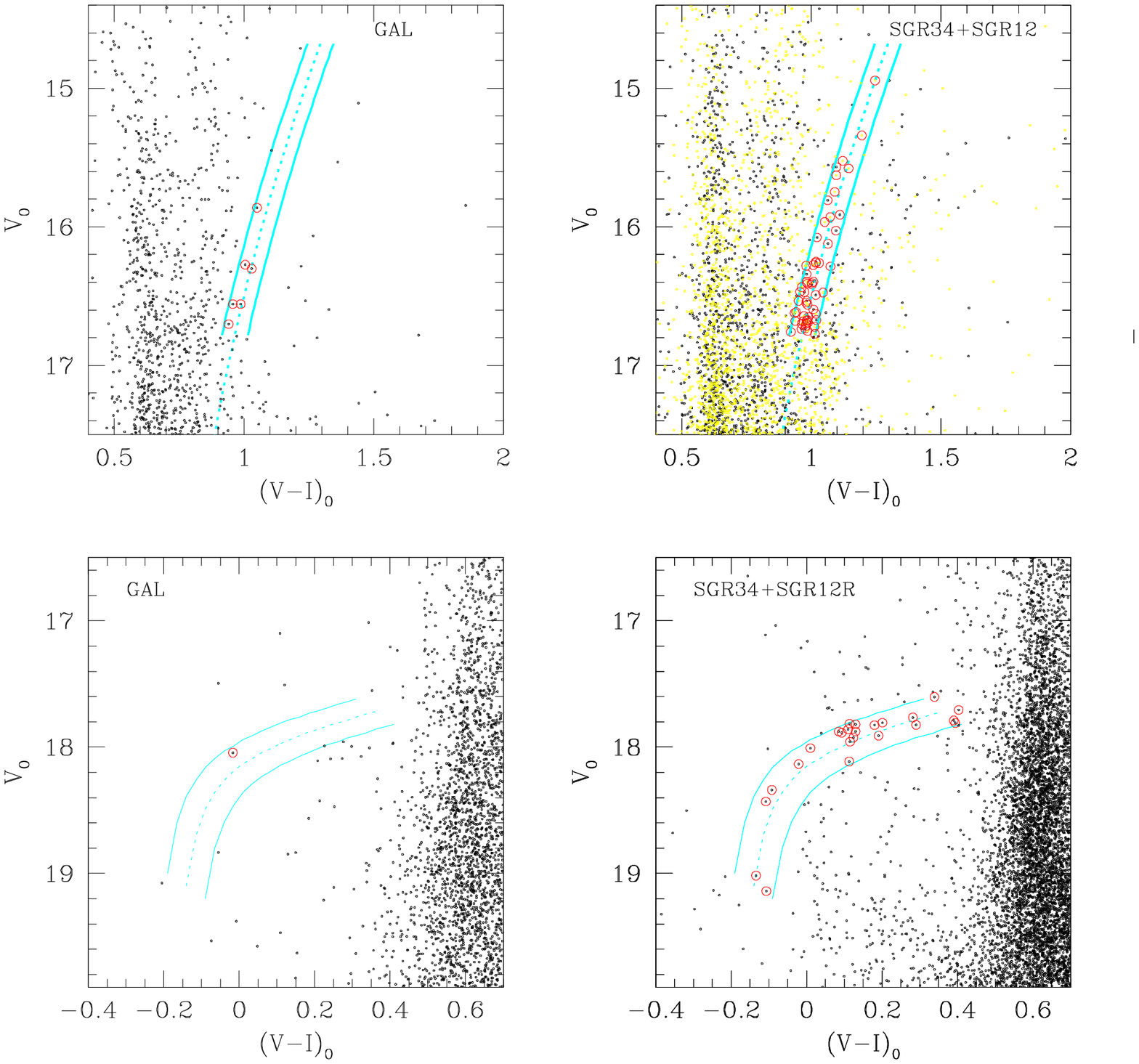}
 \caption{The RGB and HB mean ridge lines (dashed line) of Ter 8 (by MoAL)
 overplotted on the CMD of the GAL Field (left panels) and of the SGR34+SGR12
 Fields (right panels). Solid lines represent $\pm 0.05$ color edges with
 respect to the mean ridge lines. Stars lying in the strips are put in evidence
 by open circles. The samples actually adopted in each case are also indicated.}
\end{figure*} 

In fig. 15, the Ter 8 ridge line is superposed to the SGR34 CMD. 
Two main features of the Ter 8 population can be unambiguously identified
in the SDGS CMDs, clearly emerging from the sequences formed by either Sgr and
CP stars:
the brightest part of the RGB (between
the redder Sgr RGB and the CP sequences, above $V_0\sim 16.5$; hereafter T8RGB) 
and the blue HB tail around $V_0 \sim 18.2$ and between 
$-0.2 \le (V-I)_0 \le 0.4$ (hereafter T8HB). The excess of
stars around this features in the Sgr Fields CMDs, with respect to 
the control GAL sample, can be considered as a signature of the presence of a 
``Ter 8 - like'' population.

In fig. 16 the location of the Ter 8 ridge line (dashed line) is reported on
the GAL (left panels) and on the SGR34+SGR12 CMDs (right panels). 
Upper panels shows
the RGB region and lower panels show the blue HB region. The continuous lines
represent $\pm 0.05$ edges (in color), with respect to the ridge line. 
Stars lying in these strips are evidenced by open 
circles. These stars can be reasonably considered to belong to a Ter 8-like
population.

The excess of such stars in the Sgr CMDs with respect to the GAL one is
clear from the comparison between right panels and left panels in fig. 16.

However, a truly significant comparisons can be performed only after the 
correction for the different area sampled by the GAL Field 
and by SGR34+SGR12 has been applied to star counts in the defined strips
(completeness corrections are irrelevant at the considered magnitude levels).

Once the area-normalization has been performed, we can test  
the null hypothesis that {\it the difference between star 
counts in the T8RGB and T8HB regions in the SGR34+SGR12 and GAL samples be
zero}, i.e. the {\em Ter 8 - like} stars are associated to the contaminating
population and NOT to Sgr. The measured excess in star counts rules out such
hypothesis at $4.3 \sigma$ level for the blue HB stars and at $3.5 \sigma$
level for the T8RGB stars. We are aware that a number of uncertainties can
affect this result (small number statistics or errors in the distance moduli,
for instance) but a casual star count excess in two independent strips would be
a too unlucky coincidence. Furthermore, the same procedure was also applied to 
the MAL and MUSKKK CMDs and fully consistent results have been found. The
application of the test to the SGRWEST sample is a risky task because of
(a) the very high degree of foreground contamination in this Field and
(b) the relatively low number of Pop A Sgr stars, whose the searched population
is a further  subsample. Despite these drawbacks a slight Ter 8 -like star 
excess is found also in SGRWEST.       

We take the above result as {\em the first direct identification of very metal
poor stars in the \sgr galaxy}. Direct spectroscopic follow-up of a consistent
sample of T8RGB stars can confirm or rule out this claim by simultaneously
determining (i) the membership to Sgr, via radial velocity estimates and (ii)
the metal content through suitable metallicity indexes measurements.
Associating an old age (Ter 8 -like) to the detected metal
poor population seems a very reasonable extrapolation, but cannot be directly
supported by SDGS CMDs, at this stage.   

Finally by counting the number of Sgr RGB stars redder than the T8RGB strip but
confined in the same magnitude range, a rough estimate of the relative
abundance of Ter 8 -like stars in Sgr can be obtained. It turns out that, with
the above assumptions, nearly $\sim < 30 \%$ of Sgr RGB stars can be associated 
with the metal poor component, not inconsistent with the large number of 
RR Lyrae stars probably associated with Sgr 
(MKSKKU, Alard et al. 1996, Alcock et al. 1996, Mateo et al. 1996, 
Lepri et al. 1997). Note that candidate RR Lyraes has also been found in 
Ter 8 (MoAL). 

We hope that statistical decontamination on CMDs can
help providing stronger constraints on this subject, but the small number of
involved stars is not promising, in this sense.

\section{Summary and Conclusions}
      
We have described the main characteristics of a large photometric survey (SDGS)
devoted to the study of stellar populations in the \sgr galaxy. The power of
the SDGS approach is to couple deep and accurate CCD photometry of large
samples to a wide spatial baseline. 

The analysis
presented in this paper (Pap I) was aimed to establish the fundamental 
properties of the collected samples as internal homogeneity within Fields, 
the amount of interstellar reddening, completeness etc., 
that will be at the base of future detailed study of the statistically 
decontaminated SDGS CMDs (Pap II). Pap II will primarily deal with age
and metallicity of the Sgr stellar populations. 

A preliminary analysis of the most evident CMD features has also been carried
on. Furthermore we performed many test performing a local decontamination from
foreground star contributions in small regions of the CMDs. 
In our view, the adopted technique is complementary to statistical 
decontamination of the whole CMD (Pap II). The latter is the ideal tool to
study the {\em shape} of the observed sequences (i.e. for the unambiguous
identification of the TO point and to derive safe ridge lines, for example), 
while the former is more useful to
estimate the degree of contamination in the various fields and to 
put in evidence the main differences in the spatial distribution of stellar
components.  

In the following, the most interesting results are shortly described and 
discussed.

\begin{itemize}

\item {\bf Metallicity spread and gradient:} 
the observed color spread of Sgr RGB stars 
indicate a wide range of metal content, from $[Fe/H]\sim -1.5$ to 
$[Fe/H]\sim -0.7$. At present, this is the stronger element witnessing the
presence of a mix of stellar populations in Sgr Pop A.

Both RGB and HB morphology suggest the presence of a more metal rich component
in the central region of the galaxy (SGR12), confirming previous claims about
a metallicity ``radial'' gradient in Sgr (SL95, MAL). An association of the
very red HB present in the SGR12 field with a young population 
[instead of (or together with) one with a higher metal content] 
cannot be excluded at the present stage.

\item {\bf Detection of a very metal poor population:} through comparison of
SDGS CMDs with the ridge line of the CMD of the old and metal poor Sgr globular
Ter 8, an excess of stars similar to those of this cluster has been found.
These stars have been identified as the Sgr ``field'' counterpart of the
ancient component of the Sgr globular cluster system (see MoAL). The only
previous hint for the presence of such a population in the Sgr stellar mix was
due to the detection of RR Lyrae stars in the galaxy.

\item {\bf Density distribution asymmetry along the major axis:} 

The SGRWEST region (nearer to the Galactic bulge and
the leading head of Sgr along its orbit, according to IGWIS) seems significantly
deficient of Sgr stars with respect to SGR34. 

If this result is exact, it would
be very difficult to find a physical mechanism other than tidal interaction
with the Galaxy that can produce such an asymmetry in the density distribution
of Sgr stars, given also the remarkable homogeneity of the stellar content.
So, it can be considered as a clue suggesting that the tidal disruption of the
Sgr galaxy is currently going on.   

It can be conceived that Sgr stars in the
SGRWEST region are unbound from the parent galaxy and are slowly spreading
along the Sgr orbital path (Johnston 1998, Irwin 1998). 

\item {\bf Pop A and Pop B stars have similar distributions:} if the stars in
the Blue Plume identified in the CMDs are associated with a younger population
(Pop B) it must be concluded that, or {\em (a)} both episodes of star formation
(Pop A and Pop B) occurred on large scales, comparable with the dimension of the
galaxy itself (i.e. stars were formed everywhere nearly at the same epoch), or
{\em (b)} a very efficient mechanism for star mixing has been at work in Sgr.
The two-body relaxation time for this system is, as expected, much greater 
than one Hubble time for any reasonable assumption about the structure of the
galaxy, so hypothesis {\em (a)} seems the most likely.

\end{itemize}

\section*{Acknowledgments}
 
This paper is based on data taken at the New Technology Telescope (ESO, La
Silla, Chile) in the first two nights after the {\em NTT Big Bang}.We are
grateful to the ESO-NTT staff for their help and assistance during the
observation run.
     
Much of the data analysis has been made easier by the computer codes developed
at the {\em Osservatorio Astronomico di Bologna} by Paolo Montegriffo.
Barbara Paltrinieri is acknowledged for her kind assistance during the data
reduction phase. We thank Flavio Fusi Pecci, Livia Origlia and Monica Tosi for 
many useful discussions.

The financial support of the {\it Ministero delle Universit\`a e della
Ricerca Scientifica e Tecnologica} (MURST) and of the {\it Agenzia
Spaziale Italiana} (ASI) is kindly acknowledged.

This research has made use of NASA's Astrophysics Data System Abstract Service.
%
%


\begin{thebibliography}{99}
\bibitem{al} Alcock C. et al., 1997, ApJ, 474, 217
\bibitem{ard} Alard C., 1996, ApJ, 458, L17
\bibitem{bah1} Bahcall J.N., Soneira R.M., 1980, ApJS, 44, 73
\bibitem{bah2} Bahcall J.N., Soneira R.M., 1984, ApJS, 55, 67
\bibitem{bu83} Buonanno R., Buscema G., Corsi C.E., Ferraro I., Iannicola
G., 1983, A\&A, 126, 278
\bibitem {bh} Burstein D., Heiles C., 1982, AJ, 87, 1165
\bibitem{dca90} Da Costa G.S., Armandroff T.E., 1990, AJ, 100, 162 (DCA90)
\bibitem{dca95} Da Costa G.S., Armandroff T.E., 1995, AJ, 109, 2533 
\bibitem{fah} Fahlman G.G, Mandushev G., Richer H.B., Thompson I.B.,
Sivaramakrishnan A., 1996, ApJ, 459, L65
\bibitem{fer} Ferraro F.R., Clementini G., Fusi Pecci F., Buonanno R., 
Alcaino G., 1990, A\&AS, 84, 59
\bibitem{grill} Grillmair C.J., et al., 1998, AJ, 115, 144
\bibitem{hegi} Hernandez X., Gilmore G., 1998, MNRAS, in press
(astro-ph/9802261)
\bibitem{ige} Ibata R.A., Geraint F., 1998, ApJ, 500, 575 
\bibitem{igi1} Ibata R.A., Gilmore G., Irwin M.J., 1994, Nature, 370, 194
(IGI-I)
\bibitem{igi2} Ibata R.A., Gilmore G., Irwin M.J., 1995, MNRAS, 277, 781
(IGI-II)
\bibitem{b5} Ibata R.A., Wyse R.F.G., Gilmore G., Irwin M.J.,
Suntzeff N.B., 1997, AJ, 113, 634 (IGWIS)
\bibitem{ir} Irwin, M.J., 1998, in The Stellar Content of the Local Group, IAU
Symp. 192, R. Cannon \& P. Whitelock eds., S. Francisco: ASP, in press
\bibitem{jo} Johnston K.V., 1998, ApJ, 495, 297
\bibitem{hu} Hunter D.A., 1998, in T. Richtler and J.M. Braun, Eds., The 
Magellanic Clouds and Other Dwarf Galaxies. in press 
\bibitem{ko} Koribalski B., Johnston S., Optrupceck R., 1995, MNRAS, 270, L43
\bibitem{lan} Landolt A.U., 1993, AJ, 104, 340
\bibitem{ls} Layden A. C., Sarajedini A. 1997, ApJ, 486, L107 (LS97)
\bibitem{le} Lepri S., Mateo M., Layden A., Lemley S., Olzewski E.,
Morrison H., 1997, BAAS, 191.8102
\bibitem{cla} Maraston C., 1998, MNRAS, submitted
\bibitem{m97} Marconi G., Buonanno R., Castellani M., Iannicola G.,
Pasquini L., Molaro P., A\&A, 330, 453 (MAL)
\bibitem{ma1} Mateo M., Udalski A., Szymansky M., Kaluzny J., Kubiak M.,
Krzeminski W., 1995a, AJ, 110, 1141 (MUSKKK)
\bibitem{ma3} Mateo M., Kubiak M., Szymanski M., Kaluzny J., 
Krzeminski W., Udalski A., 1995b, AJ, 110, 1141 (MKSKKU)  
\bibitem{ma2} Mateo M., Mirabal N., Udalski A., Szymanski M., Kaluzny J.,
Kubiak M., Krzeminski W., Stanek K.Z., 1996, ApJ, 458, L13
\bibitem{migh} Mighell K., Armandroff T., Sarajedini A., Layden A., Mateo
M., Fusi Pecci F., Ferraro F., Buonanno R., 1997, BAAS, 190, 35.05
\bibitem{MoAL} Montegriffo P., Bellazzini M., Ferraro F.R., Martins D.,
Sarajedini A., Fusi Pecci F., 1998, MNRAS, 294, 315 (MoAL)
\bibitem{fu} Fusi Pecci F., Bellazzini M., Cacciari C., Ferraro F.R., 1995, AJ,
110, 1664
\bibitem{oa} Olszewski E.W., Aaronson M., 1985, AJ, 90, 2221
\bibitem{renz} Renzini A., 1998, AJ, in press (astro-ph/9802186)
\bibitem{rile} Rieke G.H., Lebovski M.J., 1985, ApJ, 288, 618 (RL) 
\bibitem{sa} Sarajedini A., 1994, AJ, 107, 618
\bibitem{sl1} Sarajedini A., Layden A., 1995, AJ, 109, 1086 (SL95)
\bibitem{sl2} Sarajedini A., Layden A., 1997, AJ, 113, 264 (SL97)
\bibitem{sh1} Schlegel D.J., Finkbeiner D.P., Davis M., 1997, BAAS, 191,
87.04
\bibitem{sh2} Schlegel D.J., Finkbeiner D.P., Davis M., 1998, ApJ, in press
\bibitem{si} Siegel M.H., Majewski S.R., Reid I.N., Thompson I., Landolt
A.U., Kunkel W.E., 1997, BAAS, 191.8103
\bibitem {sh} Smecker-Hane T. A., Stetson P. B., Hesser J. E., 
Lehnert M. D. 1994, AJ, 108, 507
\bibitem{sh2} Smecker-Hane T. A., Mc William A., Ibata R.A., 1998, BAAS,
192, 66.13
\bibitem{stan}Stanek K.Z., 1998, ApJ L, submitted (astro-ph/9802093)
\bibitem{to} Tosi M., 1993, in P. Prugniel and G. Meylan, Eds., ESO/OHP 
Workshop on Dwarf Galaxies. ESO, Garching, p. 143
\bibitem{wa} Walsh J.R., Dudziak G., Minniti D., Zjilstra A.A., 1997, ApJ,
487, 651
\bibitem{w} Whitelock P.A., Irwin M., Catchpole R.M., 1996, New Ast., 1, 57
(WIC)
\bibitem{za} Zhao H., 1998, ApJ, 500, L149
\bibitem{zj} Zjilstra A.A., Walsh J.R., 1996, A\&A 312, 21
\bibitem{zj1} Zjilstra A.A., Giraud E., Melnick H., Dekker H., D'Odorico
S., 1996, ESO Operating Manual n. 15, Version n. 3.0
\end{thebibliography}
\end{document}